\documentclass[aps,prx,floats,floatfix, showpacs,superscriptaddress,nofootinbib,twocolumn]{revtex4}
\usepackage{latexsym,amsmath}
\usepackage{amsbsy}
\usepackage[usenames]{color}
\usepackage[pdftex]{graphicx}
\usepackage{amsfonts,amsmath,amssymb,hyperref}
\usepackage{epstopdf}
\usepackage[tight]{subfigure}
\usepackage{color}
\usepackage{hyperref}
\usepackage{todonotes}
\usepackage{algorithm}
\usepackage{algorithmic}
\usepackage[usenames]{color}

\usepackage{bbm}
\usepackage{array}
\usepackage{multirow}

\begin{document}
\title{Latent geometry of bipartite networks}

\author{Maksim Kitsak}
\affiliation{Department of Physics, Northeastern
University, 110 Forsyth Street, 111 Dana Research Center, Boston, MA 02115, USA.}
\author{Fragkiskos Papadopoulos}
\affiliation{Department of Electrical Engineering, Computer Engineering and Informatics, Cyprus University of Technology, 33
Saripolou Street, 3036 Limassol, Cyprus}
\author{Dmitri Krioukov}
\affiliation{Department of Physics, Department of Mathematics, Department of Electrical\&Computer Engineering,
Northeastern University, 110 Forsyth Street, 111 Dana Research Center, Boston, MA 02115, USA.}

\date{\today}


\begin{abstract}
Despite the abundance of bipartite networked systems, their organizing principles are less studied, compared to unipartite networks. Bipartite networks are often analyzed after projecting them onto one of the two sets of nodes. As a result of the projection, nodes of the same set are linked together if they have at least one neighbor in common in the bipartite network. Even though these projections allow one to study bipartite networks using tools developed for unipartite networks, one-mode projections lead to significant loss of information and artificial inflation of the projected network with fully connected subgraphs. Here we pursue a different approach for analyzing bipartite systems that is based on the observation that such systems have a latent metric structure: network nodes are points in a latent metric space, while connections are more likely to form between nodes separated by shorter distances. This approach has been developed for unipartite networks, and relatively little is known about its applicability to bipartite systems. Here, we fully analyze a simple latent-geometric model of bipartite networks, and show that this model explains the peculiar structural properties of many real bipartite systems, including the distributions of common neighbors and bipartite clustering. We also analyze the geometric information loss in one-mode projections in this model, and propose an efficient method to infer the latent pairwise distances between nodes. Uncovering the latent geometry underlying real bipartite networks can find applications in diverse domains, ranging from constructing efficient recommender systems to understanding cell metabolism.
\end{abstract}

\maketitle

\section{Introduction and Motivation}
\label{sec:intro}

Many real-world networks have a bipartite structure: nodes can be separated into two disjoint sets and links exist only between nodes of different sets. Real bipartite networks are often characterized by two common properties: (i) heterogeneity in distributions of node degrees; and (ii) a large number of common neighbors shared between pairs of nodes. The heterogeneity of degree distributions has been studied extensively in both traditional (unipartite) and bipartite networks and comes as no surprise.  In many bipartite systems heterogeneous degree distributions can be approximated by power laws, $P(k)\sim k^{-\gamma}$, which are observed for at least one of the two sets of nodes~\cite{Newman2010,latapy2008basic,Tian2012,Zhang2013,diaz2010competition,nacher2009mathematical}. The second property, on the other hand, has not been studied extensively and deserves a thorough investigation.

\begin{figure}
\includegraphics[width=3in]{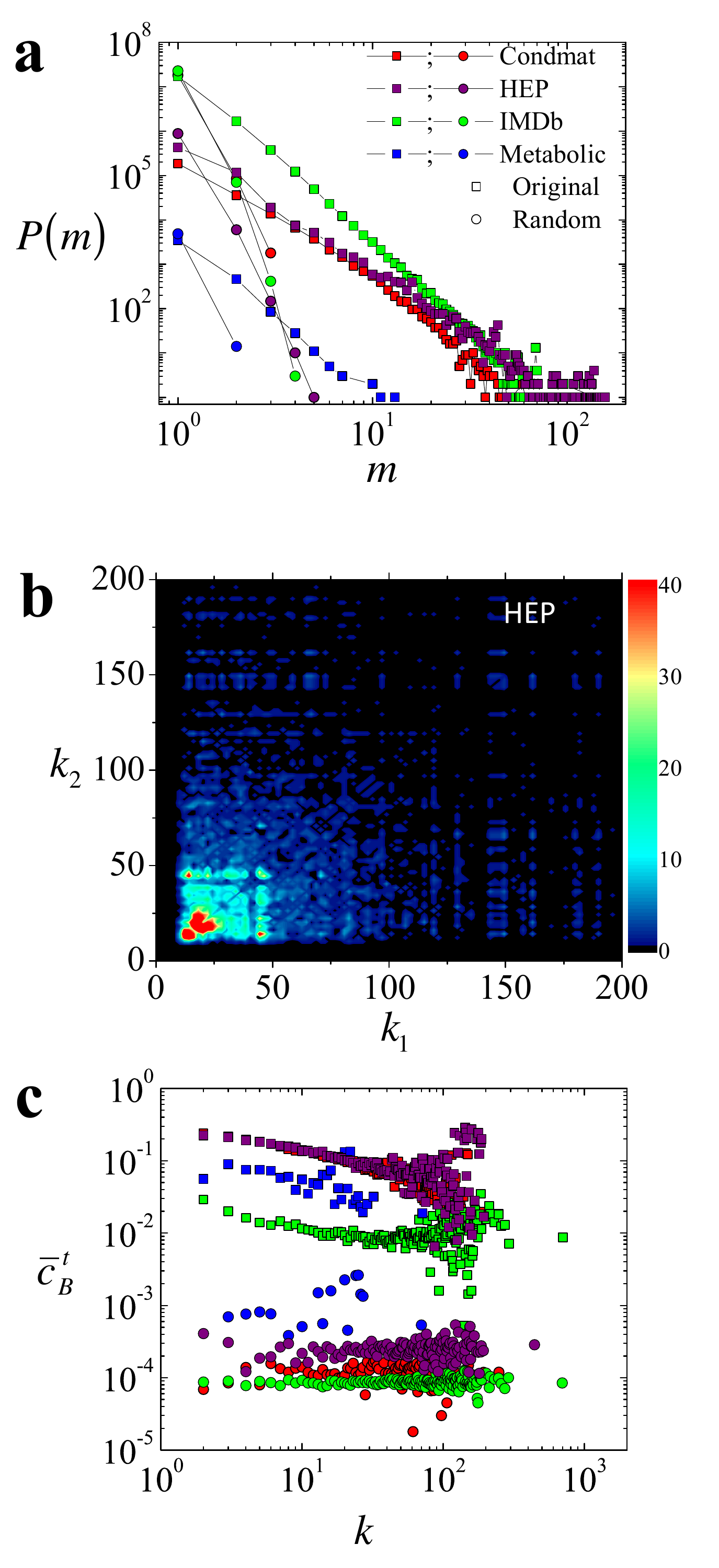}
\caption{ \footnotesize (color online) Number of common neighbors and clustering in real bipartite networks, and in their degree-preserving randomized counterparts. {\bf a}, The distribution of the number of common neighbors, $P(m)$, for pairs of top nodes in the Condmat and HEP collaboration networks, the IMDb network, and the metabolic network. The results for the original real networks are shown with squares while for their degree-preserving randomized counterparts are shown with circles. By \emph{top} nodes in each network we refer respectively to authors, actors and metabolites, while manuscripts, films and reactions are referred to as \emph{bottom} nodes (Appendix~\ref{app:data}).
 {\bf b}, Number of HEP author pairs having $m \geq 10$ common publications as a function of their total number of publications $k_{1}$ and $k_{2}$. {\bf c}, Average bipartite clustering coefficient of top nodes, $\overline{c}_{B}^{t}$, as a function of their degree $k$ for the same real networks (squares) as in panel {\bf a}, and for their degree-preserving randomized counterparts (circles). The bipartite clustering coefficient is significantly larger in the original real networks  compared to their randomized counterparts.}
\label{fig:real_clustering}
\end{figure}

In Fig.~\ref{fig:real_clustering}{\bf a} we show the distribution $P(m)$ of the number $m$ of common neighbors between nodes in several real bipartite systems: the actor-film network derived from the International Movie Database (IMDb)~\cite{imdb_online}, condensed matter (Condmat) and high energy physics (HEP) collaboration networks derived from the arXiv~\cite{arxiv}, Wikipedia~\cite{wikipedia}, and the network of metabolic reactions~\cite{ma2003reconstruction}~(see Appendix~\ref{app:data}). For each of these networks, we calculated the distribution of the number of common neighbors shared between pairs of nodes, $P(m)$. We see that the number of common neighbors in these networks is power-law distributed,
\begin{equation}
\label{eq:pm}
P(m) \sim m^{-\tau},~\tau > 2,
\end{equation}
so that significant fractions of node pairs share many common neighbors.
Similar observations have been made for other bipartite systems. For instance, the probability of two insect species pollinating $m$ different kinds of flowers in common has been shown to follow a truncated power law~\cite{burgos2008two}. Similarly, a fat-tail distribution of the number of shared requests between two users has been observed in peer-to-peer networks~\cite{iamnitchi2004small}.

To better understand the mechanisms leading to the abundance of common neighbors in bipartite systems we first ask if the observed fat-tail distributions of $P(m)$ are the consequence of heterogeneous degree distributions $P(k)$. To answer this question we randomly rewire our real bipartite networks by preserving the degrees of individual nodes~(see Appendix~\ref{app:degree_randomization}). We find that $P(m)$ in the randomized networks exhibits very fast decays, such that the maximum number of common neighbors between nodes is very small, see Fig.~\ref{fig:real_clustering}{\bf a}. This result suggests that the heterogeneity of $P(m)$ is not caused by the heterogeneity of $P(k)$. Second, we also check if the heterogeneous shape of $P(m)$ is driven by all pairs of nodes in the network or by a handful of high degree nodes. To this end, we focus on node pairs with a large number of common neighbors. We create a heatmap by counting pairs of HEP authors sharing at least $m=10$ publications, and whose publication record sizes, i.e., degrees, are $k_{1}$ and $k_{2}$. As seen from Fig.~\ref{fig:real_clustering}{\bf b}, author pairs with $m \geq 10$ common publications do not necessarily consist of authors that have published a large number of papers, as one would expect from a random collaboration pattern. On the contrary, we see that the majority of author pairs with at least $10$ common publications involve authors who barely published over $10$ publications each. This observation is not specific to the HEP collaboration network: we checked that in all considered networks both small and large degree nodes can have a large number of common neighbors.

The heterogeneity in the observed number of common neighbors implies the existence of a large number of $4$-loops in real bipartite networks. Indeed, a pair of nodes $a, b$ sharing a large number of common neighbors will have a large number of $4$-loops passing through them, that is, loops of the form $a \to c \to b \to d \to a$. Supporting this observation, we also find that real bipartite systems are characterized by strong bipartite clustering, which quantifies the density of $4$-loops in the network, see the definition in Sec.~\ref{sec:bipartite_clustering}. Bipartite clustering is typically several orders of magnitude larger in the original networks compared to their degree-preserving randomized counterparts, Fig.~\ref{fig:real_clustering}{\bf c}. We also note that similar clustering-related heterogeneity has also been observed in unipartite networks. Many real unipartite networks have been shown to exhibit power-law distributions of edge multiplicity, defined as the number of triangles shared by edges~\cite{zlatic2012networks}.

\begin{figure}
\includegraphics[width=3in]{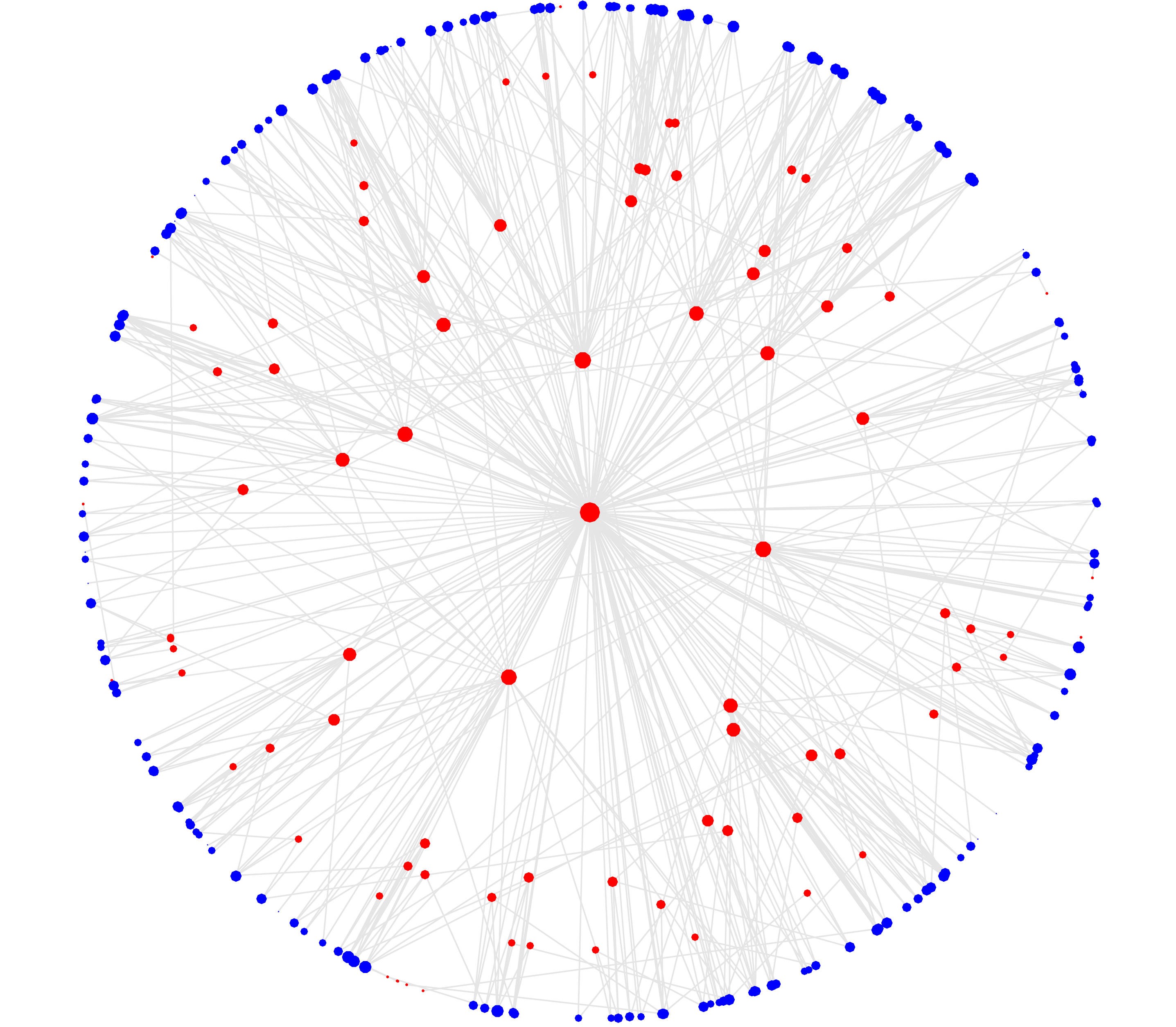}
\caption{\footnotesize (color online) Schematic illustration of a synthetic bipartite network in latent geometric space.  The network consists of $N=100$ top (red) nodes and $M=200$ bottom (blue) nodes, connected as described in Section~\ref{sec:the_model}. The nodes are placed on a disc of radius~$1$. Their angular coordinates are random on $[0,2\pi]$. The radial coordinates of all bottom nodes are $r_{j}=1$,~$\forall j$, so that they all are at the disc edge, while the radial coordinates of top nodes are $r_{i} = 1 - { \frac {\ln k_i } {\ln k_{\rm max}}}$, where  $k_i$ is the degree of node $i$ and $k_{\rm max}$ is the largest node degree of the top domain. Node sizes are proportional to the logarithm of node degrees.}
\label{fig:illustr}
\end{figure}

\subsection{Latent geometry of bipartite networks}
Here we show that the observed common properties of real bipartite networks can be explained by the existence of latent geometric spaces underlying these networks. That is, we assume that nodes in  bipartite networks are  points in some geometric space underlying the system. The latent coordinates of nodes in the space abstract node attributes, while latent distances between nodes play the role of a generalized similarity measure: the more similar the two nodes, the smaller the latent distance between them, the higher the probability that the nodes are connected, Fig.~\ref{fig:illustr}. To illustrate, consider the IMDb network for instance, where actors are linked to films they starred in. Clearly, connections in this network are not random. Both actors and films can be characterized by numerous attributes, so that connections are the result of a certain mutual match between these attributes. These attributes include genres, geographic locations of films, film release dates, etc. Similar parallels can be drawn for other systems. For instance, connections between authors and manuscripts in collaboration networks are driven by many factors, including the expertise of the authors, their geographic location, their methodology, and so on. Collectively these attributes define similarity distances between nodes in a latent space.

The large numbers of common neighbors and strong clustering observed in real bipartite systems are then a reflection of the triangle inequality in the latent space. Consider, for example, two actors $A_1$ and $A_2$ and a film $F$ mapped to three points in a metric space. The triangle inequality in the space prescribes that $d(A_2,F) < d(A_1,A_2) + d(A_1,F)$, where $d(X,Y)$ denotes the distance between nodes $X,Y$ in the space. If distances $d(A_1,A_2)$ and $d(A_1,F)$ are small, then $d(A_2,F)$ is also small, so that both actors $A_1, A_2$ are likely to co-star in film $F$, that is, $F$ is likely to be a common neighbor of $A_1$ and $A_2$. As the number of common films between two actors $A_1, A_2$ increases, so does bipartite clustering, that is, the number of loops of the form $A_1 \to F_1 \to A_2 \to F_2 \to A_1$. We stress the importance of the metric property in a latent space: if latent distances do not satisfy the triangle inequality, then bipartite networks built using these distances do not have many common neighbors and strong bipartite clustering~(Appendix~\ref{app:non_metric_models}).

\subsection{Organization of the manuscript}
To further support our geometric assumption we organize the rest of the manuscript as follows.

We begin with the review of related work in Section~\ref{sec:previous_work}. In Section~\ref{sec:the_model} we conduct a detailed analysis of a model of synthetic bipartite networks constructed in the latent space. We show that the model generates bipartite networks with either heterogeneous or homogeneous degree distributions in a given node domain, and with power-law distributions of the number of common neighbors and strong bipartite clustering. In Section~\ref{sec:one_mode_projections}, we investigate how the latent geometry of bipartite systems is transformed under one-mode projections, and prove that one-mode projections can not fully preserve latent geometry. However, we also show that under certain conditions latent geometry can be preserved approximately. In Section~\ref{sec:inference}, we propose a procedure for efficient estimation of the latent pairwise distances between pairs of nodes. Final remarks are in Section~\ref{sec:conclusion}.

\section{Related work}
\label{sec:previous_work}

Bipartite networks have been successfully used to model a large array of complex systems including collaboration networks~\cite{moody2004structure,tomassini2007empirical}, metabolic reactions~\cite{palsson2015systems,ma2003reconstruction,smart2008cascading}, peer-to-peer networks~\cite{iamnitchi2004small,liu2007building}, and recommender systems~\cite{zhou2007bipartite,lu2012recommender,bobadilla2013recommender,schafer1999recommender}. Bipartite networks can be represented as hypergraphs, generalizations of  graphs where a single edge can connect multiple nodes~\cite{berge1973graphs}. Hypergraphs, in their turn, are further generalizable to multipartite hypergraphs, where hyperedges may connect several nodes of different type. Recently, tripartite hypergraphs have been  proposed to model tagged social networks, also known as folksonomies~\cite{ghoshal2009random,liu2011self,tibely2013extracting,ming2009diffusion,zhang2010hypergraph}.

The concept of latent space has been initially introduced to model homophily and similarity in social networks~\cite{McPherson2001,mcfarland1973social}. Lately,
latent space models are attracting great interest in many diverse fields including sociology~\cite{sarkar2005dynamic,boguna2004models,csimcsek2008navigating}, statistical physics~\cite{ferretti2011preferential,Barthelemy2011} and computer science~\cite{sarkar2011theoretical,ferretti2011preferential,Barthelemy2011}.
Another closely related research area is that on random geometric graphs, well-studied in mathematics and engineering~\cite{Penrose2003,Balister2008RGGs,FraMe08-book},
particularly due to its relevance to wireless networks~\cite{FraMe08-book}.

Two  models of random bipartite geometric graphs have been proposed recently. The first model is the AB random geometric graph (AB RGG)~\cite{iyer2012percolation}, defined as the two sets of points scattered as two independent Poisson processes in Euclidean space with connections between points from different sets established if distance between them is less then the threshold distance. AB RGGs are  motivated by wireless networks where transmission and reception of a signal occurs at different frequencies~\cite{tse2005fundamentals}. Thus, of primary interest in AB RGGs are the connectivity and percolation properties~\cite{iyer2012percolation,penrose2014continuum}.

The second model is inspired by the hidden variable formalism~\cite{Caldarelli2002,Boguna2003class}: network nodes map to points in the latent space and connections between them are drawn with probabilities depending on distances between the nodes in the underlying space. It was shown in~\cite{Krioukov2010,Papadopoulos2012,soft:comm} that if latent geometry is hyperbolic, then random geometric graphs in it reproduce common structural and dynamical properties of unipartite networks -- scale-free degree distribution, strong clustering, community structure, and large-scale growth dynamics. Equivalent to hyperbolic random graphs are random graphs in Euclidean space with power-law distributed hidden variables~\cite{Krioukov2010}. This model has been recently generalized to bipartite networks in~\cite{Serrano2012}, where it was called the $\mathbb{S}^{1}\times \mathbb{S}^{1}$ model, and utilized to study cell metabolism.

\section{Topological properties of bipartite networks as reflections of latent geometry}
\label{sec:the_model}

In this section we conduct a detailed analysis of the  $\mathbb{S}^{1}\times \mathbb{S}^{1}$ model of a bipartite network in the simplest compact latent space, circle $\mathbb{S}^{1}$.

\subsection{Definitions and the $\mathbb{S}^{1}\times\mathbb{S}^{1}$ model}

We refer to the two groups of nodes in a bipartite network as top and bottom nodes, and denote their number by $N$ and $M$ respectively.
Within the modeling framework we consider, network nodes map to points in a geometric space, and as a result, every node of the network is characterized by its coordinates in this space. Both top and bottom nodes belong to the same space and, thus, distances are defined between all pairs of nodes. Yet, to generate bipartite networks, connections are allowed only between nodes of different domains. To achieve heterogeneity in node degrees and to allow some nodes to connect over large distances, every node is also assigned a hidden variable. To distinguish top and bottom nodes we denote these hidden variables as $\{\kappa_{i}\}$ and $\{\lambda_{j}\}$ respectively.

To form a bipartite network, every top-bottom pair of nodes $i,j$ is connected with a probability $r_{ij}$, which depends on the distance between the nodes and their hidden variables,
\begin{equation}
\label{eq:r_ij}
r_{ij} = r\left( \frac {d_{ij}} {d_{c}(\kappa_{i}, \lambda_{j})}\right),
\end{equation}
where $r(x)$ is the connection probability function, $d_{ij}$ is the distance between the nodes in the geometric space, and $d_{c}(\kappa_{i}, \lambda_{j})$ is a characteristic distance scale, allowing one to vary the importance of small distances depending on the nodes' hidden variables. Even though any integrable function $r(x)$ can, in principle,  serve as the connection probability function, we use
\begin{equation}
\label{eq:conn_prob}
r(x) = \left( 1 + x^{\beta}\right)^{-1},~\beta \in (1,\infty).
\end{equation}

Our choice for the connection probability function is dictated by the maximum entropy principle~\cite{Krioukov2010} and formalizes one's intuition
that similar nodes are more likely to be connected than dissimilar nodes. Indeed, $r(x)$ is a decreasing function of $x$, such that $r(x) \to 0$ as $x\to \infty$ and $r(x) \to 1$ as $x\to 0^{+}$. Further, parameter $\beta$ in Eq.~(\ref{eq:conn_prob}) controls the abundance of long-distance connections: the larger the $\beta$, the less preferred the longer-distance connections.

We focus on the simplest realization of the model, where top and bottom nodes are placed uniformly at random on a $1$-dimensional Euclidean ring $\mathbb{S}^{1}$ of radius $R$, with probability density functions (pdfs) $\rho_{t}(\theta)=\rho_{b}(\phi)= \frac{1}{2 \pi}$.
 Given $R$, the densities of top and bottom nodes on the ring are $\delta_{t} = \frac{N}{ 2 \pi R} $ and $\delta_{b} = \frac{M} {2 \pi R }$.
The hidden variables of the top and bottom nodes are drawn at random from pdfs $\rho_{t}(\kappa)$ and $\rho_{b}(\lambda)$. To ease notation, in the rest of the paper we drop the indices from the top and bottom hidden variable distributions, i.e., $\rho_{t}(\kappa) \equiv \rho(\kappa)$ and $\rho_{b}(\lambda) \equiv \rho(\lambda)$. To achieve heterogeneous degree distributions, we choose the characteristic scale in Eq.~(\ref{eq:r_ij}) as $d_{c}(\kappa_{i}, \lambda_{j})=\mu \kappa_{i}\lambda_{j}$, which allows nodes with large hidden variables to connect over large distances with higher probability. Parameter $\mu > 0$  rescales  all latent distances and controls the expected degrees of the top and bottom domains. Without loss of generality, we set $R = \frac{N}{2\pi}$, which corresponds to the unit density of top nodes, i.e., $\delta_{t} = 1$. We are interested in large and sparse bipartite networks, $E \propto N \propto M \gg 1$, where $E$ is the number of edges.

Even though nodes from both domains in the model belong to the same Euclidean ring $\mathbb{S}^1$, we refer to the model as the $\mathbb{S}^1 \times \mathbb{S}^1$ model to emphasize its bipartite structure and to distinguish it from the $\mathbb{S}^1$ model for unipartite networks developed in~\cite{Serrano2008}. The model is fully specified by the number of top and bottom nodes $N$ and $M$, the connection probability function $r(x)$, and the pdfs $\rho(\kappa)$ and $\rho(\lambda)$. It can be summarized as follows:
\begin{enumerate}
\item Sample the angular coordinates of top nodes $\theta_i$, $i=1,2,\ldots,N$, uniformly at random from $[0, 2\pi]$, and their hidden variables $\kappa_{i}$, $i=1,2,\ldots,N$, from the pdf $\rho(\kappa)$;
\item Sample the angular coordinates of bottom nodes $\phi_j$, $j=1,2,\ldots,M$, uniformly at random from $[0, 2\pi]$, and their hidden variables $\lambda_{j}$, $j=1,2,\ldots,M$, from the pdf $\rho(\lambda)$;
\item Connect every top-bottom node pair with probability
\begin{eqnarray}
\label{eq:s1_r}
r(\kappa_{i}, \theta_{i}; \lambda_{j}, \phi_{j})  &=& {1
\over 1 + \left[ \frac{d\left(\theta_i ,\phi_j\right)}{d_{c}\left(\kappa_i, \lambda_j\right)} \right]^{\beta}},\\
d(\theta_i, \phi_j) &=& R \left(\pi - | \pi -|\theta_i - \phi_j||\right),\nonumber\\
d_{c}\left(\kappa_i, \lambda_j\right) &=& \mu  \kappa_i \lambda_j. \nonumber
\end{eqnarray}
\end{enumerate}

The hidden variables in the $\mathbb{S}^{1}\times \mathbb{S}^{1}$ model allow long distance connections among some nodes and are necessary to achieve heterogeneous degree distributions.  An alternative  approach to achieve heterogeneity in node degrees is to consider latent spaces of non-zero curvature. Even though both approaches are fully equivalent~~(see Appendix~\ref{app:h2h2}), we utilize the former approach as it is more convenient for calculations.

\subsection{Basic properties}
\label{sec:basic_properties}

The basic topological properties of synthetic networks constructed by the $\mathbb{S}^{1}\times\mathbb{S}^{1}$ model can be obtained in a straightforward manner. Since angular node coordinates are sampled uniformly at random from $[0,2\pi]$, the expected degree of a top node with hidden variable $\kappa$ and angular position $\theta$,  $\overline{k}(\kappa, \theta)$, is given by
\begin{equation}
\label{eq:avg} \overline{k}(\kappa, \theta) = \frac{M}{2\pi} \int {\rm d}\lambda\, \rho(\lambda) \int_{0}^{2\pi} {\rm d} \phi \,
r \left( \frac{d(\theta, \phi)}{ \mu \kappa \lambda } \right).
\end{equation}

Notice that due to the uniform angular distribution of nodes $\overline{k}(\kappa, \theta)$ is independent of $\theta$, $\overline{k}(\kappa, \theta)=\overline{k}(\kappa)$.
The evaluation of the inner integral in Eq.~(\ref{eq:avg}) leads to
\begin{eqnarray}
\label{eq:avg_k1_first_exact}
\overline{k}(\kappa) &=&\frac{\mu M}{\pi R} \int {\rm d} \lambda\, \lambda \rho(\lambda) K ~_{2}F_{1}\left(1, \frac{1}{\beta}, 1+\frac{1}{\beta}, -K^{\beta} \right),\\
K &\equiv& \frac{R \pi}{\mu \kappa \lambda},
\end{eqnarray}
where $~_{2}F_{1}$ is the hypergeometric function. For sufficiently large networks, the expected degrees for nodes with $\kappa$ values satisfying $M \gg \kappa$ can be approximated as
\begin{equation}
\label{avg_k1_first}
\overline{k}(\kappa)\approx \frac{\mu M I }{\pi R} \kappa \overline{\lambda},
\end{equation}
where
\begin{equation}
\label{eq:integral_beta}
I \equiv \int_{0}^{\infty} r(x) \,{\rm d} x = \left( \pi / \beta\right) {\rm csc} \left( \pi / \beta \right),
\end{equation}
and $\overline{\lambda} \equiv \int {\rm d} \lambda \, \lambda \rho(\lambda)$.  The expected degree of the entire top node domain, $\overline{k}$, is given by averaging  $\overline{k}(\kappa)$ over all possible $\kappa$ values,
\begin{equation}
\label{avg_k2_first}
\overline{k} = \int{\rm d} \kappa \, \rho(\kappa) \overline{k}(\kappa) \approx \frac{\mu M I}{ \pi R} \overline{\kappa} \overline{\lambda},
\end{equation}
where $\overline{\kappa} \equiv \int {\rm d} \kappa  \, \kappa \rho(\kappa)$. Since the model is defined symmetrically for top and bottom nodes, the expected degrees for the bottom domain can be obtained by swapping top and bottom node variables in Eqs.~(\ref{avg_k1_first}) and (\ref{avg_k2_first}),
\begin{eqnarray}
\label{avg_l1_first}\overline{\ell}(\lambda) & \approx & \frac{\mu N I}{\pi R} \lambda
\overline{\kappa},\\
\label{avg_l2_first}\overline{\ell} & \approx & \frac{\mu N I}{\pi R} \overline{\kappa} \overline{\lambda}.
\end{eqnarray}
It can be seen from Eqs.~(\ref{avg_k1_first}) and (\ref{avg_l1_first}) that $\mu$ is a dumb parameter in the sense that  for any particular value of $\mu$, one can always rescale the hidden variables assigned to top and bottom nodes, $\{\kappa_i\}, \{\lambda_j\}$, in order to obtain desired $\{\overline{k}_{i}(\kappa_i)\}$ and $\{\overline{\ell}_{j}(\lambda_j)\}$ values. Therefore, to simplify  notation we set
\begin{equation}
\label{eq:mu}
\mu = \frac{\pi R }{ M I \overline{\lambda}} = \frac{\pi R}{N I \overline{\kappa}}.
\end{equation}
The second equality in the above relation holds since the expected number of links in the bipartite network is $E=M\overline{\ell}=N \overline{k}$. Using Eq.~(\ref{eq:mu}), we can rewrite Eqs.~(\ref{avg_k1_first}-\ref{avg_l2_first}) as
\begin{eqnarray}
\label{avg_k1} \overline{k}(\kappa) &=& \kappa,\\
\label{avg_k2} \overline{k} &=& \overline{\kappa},\\
\label{avg_l1} \overline{\ell}(\lambda) &=& \lambda,\\
\label{avg_l2}\overline{\ell} &=& \overline{\lambda}.
\end{eqnarray}

\begin{figure}
\includegraphics[width=3in]{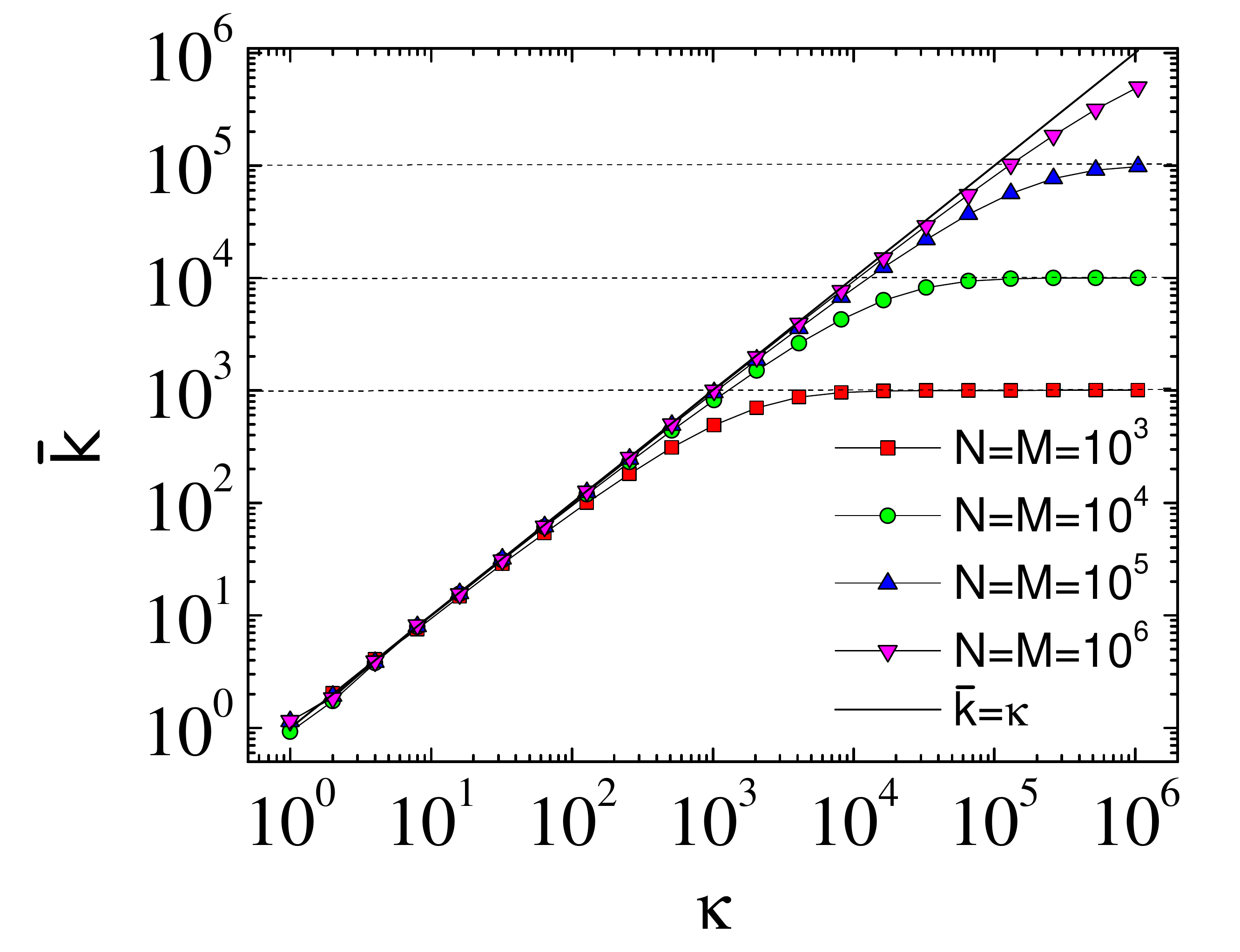}
\caption{\footnotesize (color online) Expected degree of the node, $\overline{k}$, in the $\mathbb{S}^{1}\times\mathbb{S}^{1}$ model as a function of its hidden variable $\kappa$. The $\bar{k}(\kappa)$ curves are shown  for different bottom domain sizes and $\rho(\lambda) = (\gamma_{b} - 1 )\lambda_{0}^{\gamma_{b} - 1} \lambda^{-\gamma_{b}}$ with $\gamma_{b} = 2.5$ and $\lambda_{0}$ corresponding to $\overline{\lambda}=10$. Each point is the average over $100$ realizations. Note that $\overline{k}(\kappa) \approx \kappa$ for $\kappa \ll M$, supporting Eqs.~(\ref{avg_k1_first}) and (\ref{avg_k1}). At the same time, $\overline{k}(\kappa) \to M$ as $\kappa \to \infty$. The horizontal dashed lines show corresponding values of $M$. }
\label{fig:s1s1_avg_k_approx}
\end{figure}

Eqs.~(\ref{avg_k1}) and (\ref{avg_l1}) indicate that the hidden variables of nodes are their expected degree values in the resulting topology, see Fig.~\ref{fig:s1s1_avg_k_approx}. Figure~\ref{fig:s1s1_avg_k_approx} also illustrates how the approximation in Eq.~(\ref{avg_k1_first}) becomes better for high values of $\kappa$ as the size of the network increases.

%

\subsection{Degree distributions}
\label{sec:deg_dist}

To compute the degree distributions of the top and bottom nodes we consider the propagators $g(k|\kappa,\theta)$ and $g(\ell|\lambda,\phi)$. The propagator  $g(k|\kappa,\theta)$ ($g(\ell|\lambda,\phi)$) is defined as the probability that a top (bottom) node with hidden variables $\kappa, \theta$ ($\lambda, \phi$) forms exactly $k$ ($\ell$) connections to bottom (top) nodes. Since the angular coordinates of nodes are uniformly distributed, the propagators do not depend on the node angles $\theta$ and $\phi$, and in the case of sparse bipartite networks  they can be approximated by the Poisson distribution~\cite{Kitsak2011}, that is,
\begin{eqnarray}
\label{eq:g1}g(k|\kappa,\theta) = g(k|\kappa)&\approx&
e^{-\kappa}\left[\kappa\right]^{k}/k!,\\
\label{eq:g2}g(\ell|\lambda,\phi) = g(\ell|\lambda)&\approx&
e^{-\lambda}\left[\lambda\right]^{\ell}/\ell!.
\end{eqnarray}
The degree distributions of the top and bottom domains are obtained by averaging the corresponding propagators over the possible values of the hidden variables,
\begin{eqnarray}
\label{eq:bip_pk} P(k) &=& \int {\rm d} \kappa \, \rho(\kappa) g(k|\kappa),\\
\label{eq:bip_pl} P(\ell) &=& \int {\rm d} \lambda \, \rho(\lambda) g(\ell|\lambda).
\end{eqnarray}
It can be seen from Eqs.~(\ref{eq:g1}-\ref{eq:bip_pl}) that $P(k)$ and $P(\ell)$ are independent of one another---they only depend on $\rho(\kappa)$ and $\rho(\lambda)$, respectively. Furthermore, the Poissonian character of the propagators $g(k|\kappa)$ and $g(\ell |\lambda)$ indicates that the resulting degree values of nodes in both domains are narrowly distributed around their hidden variables. This means that the functional forms of $P(k)$ and $P(\ell)$ will be similar to those of $\rho(\kappa)$ and $\rho(\lambda)$,  allowing one to construct different degree distributions by engineering proper pdfs of hidden variables $\rho(\kappa)$ and $\rho(\lambda)$. Even though real bipartite systems are characterized by different degree distributions, of our primary interest are scale-free and Poissonian distributions, which we discuss in Section~\ref{sec:categories} below.

\subsection{Degree-degree correlations}

Degree-degree correlations can be quantified using the average nearest neighbor degree (ANND), defined as the average degree of all neighbors of nodes with given degree $k$~\cite{pastor2001dynamical}. It is straightforward to verify that the $\mathbb{S}^{1} \times \mathbb{S}^{1}$ model is characterized by random degree-degree correlations due to the uniform placement of nodes on $\mathbb{S}^{1}$:
\begin{eqnarray}
\label{eq:annd1}
\overline{\ell}_{nn}(k) &=& \frac{\overline{\ell^{2}}}{\overline{\ell}},\\
\label{eq:annd2}
\overline{k}_{nn}(\ell) &=& \frac{\overline{k^{2}}}{\overline{k}}.
\end{eqnarray}
Indeed, the connection probability between two nodes with fixed hidden variables $\kappa_{i}$ and $\lambda_{j}$ is proportional to the product of these hidden variables:
\begin{equation}
r(\kappa_{i},\lambda_{j}) = \frac{1}{4 \pi^{2}} \iint {\rm d} \theta_{i}\ {\rm d} \phi_{j}\, r\left( \kappa_{i},\theta_{i};\lambda_{j} \phi_{j} \right) \propto \lambda_{i} \kappa_{j}.
\end{equation}
Then, since hidden variables are equal to expected node degrees, this result can be regarded as the soft equivalent of $p(k_{i},\ell_{j}) \propto k_{i} \ell_{j}$ in uncorrelated bipartite networks, where $p(k_{i},\ell_{j})$ is the probability that randomly chosen nodes with degrees $k_{i}$ and $\ell_{j}$ are connected. The rigorous proof of Eqs.~(\ref{eq:annd1}) and (\ref{eq:annd2}) can be obtained following the hidden variable formalism for bipartite networks~\cite{Kitsak2011}.

\subsection{Categories of bipartite networks}
\label{sec:categories}

Based on the degree distributions of the top and bottom domains, real bipartite networks often fall into two categories. The first category corresponds to networks with scale-free degree distribution in both top and bottom domains (sf/sf). The second category corresponds to networks with scale-free degree distribution in one domain and Poisson degree distribution in the other domain (sf/ps). Among the real networks that we consider, IMDb, Wikipedia, and the HEP collaboration network fall into the first category. The Condmat collaboration network and the network of metabolic reactions fall into the second category~(see Appendix~\ref{app:data}).

Since $P(k)$ and $\rho(\kappa)$ are expected to be of similar functional form, one can see that a scale-free degree distribution $P(k) \sim k^{-\gamma}$ can be obtained by using the continuous power-law distribution of $\kappa$ on $[\kappa_0,\infty)$
\begin{equation}
\rho(\kappa) = (\gamma-1) \kappa_0^{\gamma-1} \kappa^{-\gamma}
\end{equation}
with the desired (``target'') value of exponent $\gamma$ of degree distribution $P(k)$.
Indeed, it follows from Eqs.~(\ref{eq:g1}) and (\ref{eq:bip_pk}) that in this case
\begin{equation}
\label{eq:bip_pk_sf}P(k) \approx (\gamma - 1) \kappa_{0}^{\gamma-1} \frac{\Gamma\left[ k - \gamma + 1, \kappa_{0}\right]}{\Gamma\left[ k +
1\right]}\sim k^{-\gamma}.
\end{equation}
Parameter $\kappa_0>0$ is the smallest $\kappa$ value, i.e., the expected minimum node degree, which also controls the mean value of $\kappa$ \begin{equation}
\bar{\kappa}=\kappa_0(\gamma-1)/(\gamma-2),
\end{equation}
and the expected averaged degree $\bar{k}=\bar{\kappa}$~(\ref{avg_k2}).

The Poisson degree distribution can be obtained by choosing $\rho(\kappa)=\delta(\kappa-\overline{\kappa})$, where $\delta(x)$ is the Dirac delta function and $\overline{\kappa}$ is the expected degree of the domain. The degree distribution $P(k)$ in this case is given by
\begin{equation}
\label{eq:bip_pk_exp}P(k) = g(k|\overline{\kappa}) = e^{-\overline{\kappa}} \left[\overline{\kappa}\right]^{k} / k!.
\end{equation}
Due to the symmetry of the model, the degree distribution of the bottom domain can be obtained similarly
through the proper choice of $\rho(\lambda)$. The independence of $P(k)$ and $P(\ell)$ allows one to construct bipartite networks
with an arbitrary combination of degree distributions for the top and bottom domains.

For illustration, we visualize in Fig.~\ref{fig:illustr} a toy sf/ps network consisting of $N=100$ and $M=200$ nodes. The hidden variables of the top nodes are drawn from the pdf $\rho(\kappa) = (\gamma-1) \kappa^{-\gamma}$, $\gamma = 2.1$, $\kappa \in [1,\infty)$, $\bar{\kappa}=(\gamma-1)/(\gamma-2)$, while the bottom node hidden variables are chosen as $\lambda_{i} = \overline{\lambda}$ for all nodes $i$, where $\overline{\lambda}$ satisfies $N\overline{\kappa} = M \overline{\lambda}$. The connections are drawn with probability $r(x)$ prescribed by Eq.~(\ref{eq:s1_r}) with $\beta = 1.5$.

\subsection{Number of common neighbors}
\label{comn_nbrs}

\begin{figure*}
\includegraphics[width=6.5in]{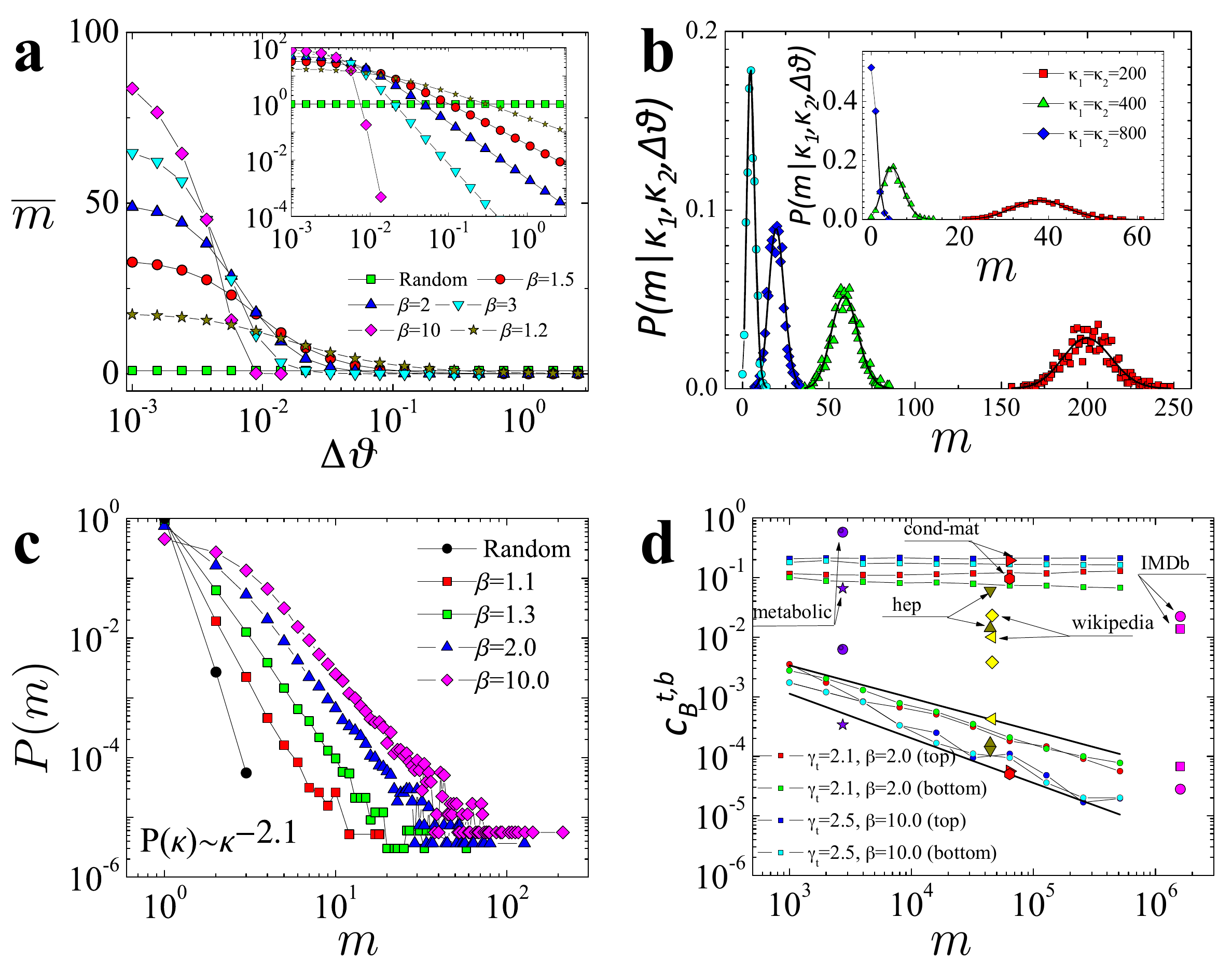}
\caption{ \footnotesize (color online) Number of common neighbors and bipartite clustering in the $\mathbb{S}^{1}\times\mathbb{S}^{1}$ model. Plots {\bf a-c}  correspond to sf/ps bipartite networks with $N=M=10^5$ nodes, $\rho(\kappa)\sim \kappa^{-2.1}$ with $\overline{\kappa}=11.0$, and $\rho(\lambda) = \delta(\lambda-1)$. The results for the  degree-preserving randomized counterparts of the networks are marked as Random. {\bf a}, The average number of common neighbors $\overline{m}$ as a function of the angular distance $\Delta\theta$ between two top nodes with $\kappa_1 = \kappa_2 = 100$, i.e., $\overline{m} \equiv \overline{m}(100, 100, \Delta\theta)$. $\overline{m}$ is calculated numerically for different values of $\beta$ using Eq.~(\ref{eq:paper_avg_m}).  Higher values of $\beta$ favor connections at smaller angular distances, i.e., between more similar top-bottom nodes. Shown in the inset, is the same plot in log-log scale. $\overline{m}$ decreases asymptotically as a power law function of $\Delta\theta$, $\overline{m} \sim \Delta\theta^{-\beta}$.
{\bf b}, Conditional distribution of the number of common neighbors, $P(m|\kappa_1, \kappa_2, \Delta\theta)$, for two top nodes with $\kappa_1 = \kappa_2 = 400$ separated by different angular distances $\Delta\theta$. From left to right: $\Delta\theta= \pi/32$, $\Delta\theta = \pi/64$, $\Delta\theta= \pi/128$, $\Delta\theta= 0$. The inset shows $P(m|\kappa_1, \kappa_2, \Delta\theta)$ for two top nodes with different values of $\kappa_1$, $\kappa_2$, separated by $\Delta\theta= \pi/32$. Solid lines are corresponding analytical results using (\ref{eq:gauss_m}).
{\bf c},  The distribution of the number of common neighbors, $P(m)$, for different values of $\beta$, calculated across all pairs of top nodes.
{\bf d}, Average bipartite clustering coefficients,  $\overline{c}^{\rm t,b}_{B}$, as a function of the network size $N$ for sf/ps modeled networks (squares) and for their degree-preserving randomized counterparts (circles). All modeled networks have the same number of top and bottom nodes, $N=M$. The network parameters $\gamma_{t}$ (power law exponent of the top domain) and $\beta$ are marked on the plot. The average degrees of the top and bottom domains are $\overline{k}=\overline{\ell}=11$ for $\gamma_{t} = 2.1$, and $\overline{k}=\overline{\ell}=3$ for $\gamma_{t} = 2.5$.  A maximum degree cut-off, $k_{\rm max} =  N^{1/2}$, was imposed in all generated networks in order to avoid structural correlations. $\overline{c}_{B}^{t,b}$  in the original networks are large and independent of their size, while in their randomized counterparts are small and vanish as $\overline{c}_{B}^{t,b}\sim N^{-\delta}$, with $\delta = \frac{\gamma_{t,b}-1 - }{2}$. The plot also shows top and bottom clustering $\overline{c}_{B}^{t,b}$ for the real bipartite networks (IMDb network, Condmat and HEP collaboration networks, Wikipedia, and the metabolic network), and for their degree-preserving randomized counterparts. We use the same symbols and colors for each real network and for its randomized counterpart, and we observe a similar behavior as in the modeled networks. Clustering in the original real networks is always orders of magnitude larger than clustering in the corresponding randomized networks.
}
\label{fig:s1s1_pm}
\end{figure*}
The number of common neighbors is the most basic non-binary network-based measure of similarity between two nodes in a bipartite network---the smaller the similarity distance between two nodes, the more similar the nodes are, and the larger is the number of neighbors they are expected to share. This makes the number of common neighbors a crucial measure, allowing one to estimate the similarity distance between two nodes in a bipartite network, Section~\ref{sec:inference}. Below, we analyze this measure in the $\mathbb{S}^{1}\times\mathbb{S}^{1}$ model.

Consider two top nodes  characterized by hidden variables $\kappa_1$ and $\kappa_2$ and angular coordinates $\theta_1$ and $\theta_2$. The probability $p_{12}$ that these two nodes are simultaneously connected to a bottom node with hidden variable $\lambda$ and angular coordinate $\phi$ is
\begin{equation}
\label{eq:p_prob}p_{12}  = r(\kappa_1, \theta_1; \lambda, \phi)r(\kappa_2, \theta_2; \lambda, \phi),
\end{equation}
where $r(\kappa, \theta; \lambda, \phi)$ is the connection probability in Eq.~(\ref{eq:s1_r}). The expected number of common neighbors between these two top nodes, $\overline{m}(\kappa_1,\theta_1; \kappa_2,\theta_2)$, can be calculated by averaging $p_{12}$ over all possible positions $\phi$ and hidden variables $\lambda$ of bottom nodes,
\begin{eqnarray}
\overline{m}(\kappa_1,\theta_1;\kappa_2,\theta_2) =\nonumber \\
\frac{M}{2\pi} \iint\rho(\lambda) r(\kappa_1, \theta_1; \lambda, \phi)r(\kappa_2, \theta_2;
\lambda, \phi) \, {\rm d} \lambda \, {\rm d} \phi.
\label{eq:paper_avg_m}
\end{eqnarray}
Due to the uniform distribution of angular coordinates, $\overline{m}(\kappa_1,\theta_1; \kappa_2,\theta_2)$ depends on the angular (similarity) distance between the two top nodes, $\Delta\theta_{12} = \pi - | \pi -|\theta_1 - \theta_2||$, and not on their individual coordinates $\theta_{1}$ and $\theta_{2}$. That is, $\overline{m}(\kappa_1,\theta_1; \kappa_2,\theta_2) \equiv \overline{m}(\kappa_1, \kappa_2, \Delta\theta_{12})$.  It is straightforward to verify (see Appendix~\ref{app:comn_nbrs_ave}) that $\overline{m}(\kappa_1, \kappa_2, \Delta\theta_{12})$ is independent of the network size $N \propto M$.  It depends only on the node hidden variables $\kappa_1, \kappa_2$ and the distance between the nodes $d_{12} = R \Delta\theta_{12} \propto N \Delta\theta_{12}$. It follows from Eq.~(\ref{eq:paper_avg_m}) that for any values of $\kappa_1, \kappa_2$, $\overline{m}(\kappa_1, \kappa_2, \Delta\theta_{12})$ decreases as the angular distance $\Delta\theta_{12}$ increases, for both domains of sf/sf and of sf/ps bipartite networks, following a power law,
\begin{equation}
\label{eq:avg_m_vs_dtheta}
\overline{m}(\kappa_1, \kappa_2, \Delta\theta_{12}) \sim \Delta \theta_{12}^{-\beta},
\end{equation}
where the exponent $\beta > 1$ is the parameter in the connection probability function in Eq.~(\ref{eq:s1_r}), see Fig.~\ref{fig:s1s1_pm}{\bf a} and Eq.~(\ref{eq:app:ru}) in Appendix~\ref{app:proj}.  The conditional probability for two nodes with hidden variables $\kappa_1, \kappa_2$ separated by angular distance $\Delta\theta_{12}$ to have $m$ common neighbors, is narrowly distributed around its ensemble average $\overline{m}(\kappa_1, \kappa_2, \Delta\theta_{12})$, and in the case of sparse bipartite networks can be approximated by the Poisson distribution
\begin{eqnarray}
\label{eq:gauss_m}
\nonumber P(m|1,2) \approx e^{-\overline{m}(\kappa_1, \kappa_2, \Delta\theta_{12})} \left[\overline{m}(\kappa_1, \kappa_2, \Delta\theta_{12})\right]^{m}/m!,\\
\end{eqnarray}
where $P(m|1,2)$ is the shorthand notation for $P(m|\kappa_1, \kappa_2, \Delta\theta_{12})$, see Fig.~\ref{fig:s1s1_pm}{\bf b} and Appendix~\ref{app:comn_nbrs_dist}.

Finally, the unconditional distribution of the number of common neighbors, $P(m)$, is obtained by averaging $P(m|1,2)$ over all possible hidden variables $\kappa_1, \kappa_2$ and angular distances $\Delta \theta_{12}$,
\begin{equation}
 P(m) = \frac{1}{\pi} \iiint   P(m|1,2)  \rho(\kappa_1)  \rho(\kappa_2) \,{\rm d} \kappa_1 \, {\rm d} \kappa_2
 \, {\rm d} \Delta \theta_{12}.
 \label{eq:pm_integral}
\end{equation}
As before, the corresponding expressions for the bottom domain nodes can be obtained by swapping the variables ($\kappa, \theta$) with ($\lambda, \phi$) and following the same analysis. The solution of the integral in Eq.~(\ref{eq:pm_integral}) depends on the functional form of $P(m|1,2)$, which in turn depends on the pdfs of the hidden variables and on the value of parameter $\beta$. While in general there is no closed-form solution to Eq.~(\ref{eq:pm_integral}),  different closed-form solutions can be obtained for integer values of $\beta$. For instance, when $\beta = 2$ and $\rho(\kappa)\sim\kappa^{-\gamma}, \rho(\lambda)=\delta(\lambda - \overline{\lambda})$ (sf/ps networks), we can show that $P(m)$ for the top domain scales as
\begin{equation}
\label{eq:m_power_law}
P(m)\sim m^{-\tau},
\end{equation}
with $\tau = 2\gamma - 3/2$~(see Appendix~\ref{app:comn_nbrs_dist}). Our numerical experiments indicate that a similar power-law scaling of $P(m)$ also holds for a range of $\beta$ values and for both domains of sf/sf networks, cf. Fig.~\ref{fig:s1s1_pm}{\bf c} and Fig.~\ref{fig:si_pm}{\bf c-h} in Appendix~\ref{app:comn_nbrs_dist}.

The power-law scaling of $P(m)$ means that a large number of node pairs have many common neighbors, and therefore, many $4$-loops passing through them, which as explained in Sec.~\ref{sec:intro} implies strong bipartite clustering. We focus on bipartite clustering below.

\subsection{Bipartite clustering}
\label{sec:bipartite_clustering}

To quantify bipartite clustering, we consider the bipartite clustering coefficient ${c}_{B}(i)$ introduced by Zhang et al.~\cite{zhang2008clustering}, which aims at quantifying the density of $4$-loops adjacent to a node $i$,
\begin{equation}
\label{eq:bip_clust}
{c}_{B}(i) = \frac{\sum_{j \neq l} \left(m_{jl} - 1 \right)}{\sum_{j \neq l} \left[ k_j + k_l - m_{jl} -1
\right]}.
\end{equation}
The summation $\sum_{j \neq l}$ goes over all pairs of neighbors $j, l$ of node $i$, $m_{jl}$ is the number of common
neighbors between $j$ and $l$, and $k_{j}$ and $k_{l}$  are the degrees of $j$ and $l$. As seen from Eq.~(\ref{eq:bip_clust}), ${c}_{B}(i)$ is essentially a normalized measure of the density of common neighbors in the vicinity of node $i$.

${c}_{B}(i)$ has also the following simple and intuitive similarity-based interpretation. Let $A_{j}$ and $A_{l}$ be the sets of neighbors of nodes $j$ and $l$ excluding node $i$. Then $m_{jl}-1$ is the size of the intersection of $A_{j}$ and $A_{l}$, $m_{jl}-1= \|A_{j} \bigcap A_{l}\|$, while $k_j + k_l - m_{jl} -1 = \|A_{j} \bigcup A_{l}\|$ is their union.  Therefore, Eq.~(\ref{eq:bip_clust}) can be written as
\begin{equation}
\label{eq:combined_jaccard}
c_B(i) = \frac{\sum_{j \neq l} \|A_j \bigcap A_l \|}{ \sum_{j \neq l} \|A_j \bigcup A_l\|}.
\end{equation}
The ratio of the intersection and union of two sets is known as the Jaccard similarity coefficient~\cite{jaccard1901etude}. $c_B(i)$ is given by the ratio of the sums of intersections and unions for all pairs of $i$'s neighbors (Eq.~(\ref{eq:combined_jaccard})). Therefore, $c_B(i)$ can be interpreted as a combined or effective Jaccard similarity of $i$'s neighbors.

The average bipartite clustering coefficients $\overline{c}^{\rm t}_{B}, \overline{c}^{\rm b}_{B}$ for the top and bottom node domains can be written as
\begin{eqnarray}
\label{eq:bip_clust1}\overline{c}^{\rm t}_{B} &=& \frac{1}{N} \sum_{i \in \Omega_{\rm t}} c_B(i),\\
\label{eq:bip_clust2}\overline{c}^{\rm b}_{B} &=& \frac{1}{M} \sum_{j \in \Omega_{\rm b}} c_B(j),
\end{eqnarray}
where $\Omega_{\rm t}$ and $\Omega_{\rm b}$ are the sets of all top and bottom nodes. Expressions for the expected bipartite clustering coefficients in the $\mathbb{S}^{1}\times\mathbb{S}^{1}$ model are derived in Appendix~\ref{app:clustering}. Qualitatively, $\overline{c}^{\rm t}_{B}, \overline{c}^{\rm b}_{B}$ are large in the $\mathbb{S}^{1}\times\mathbb{S}^{1}$ model and independent of the number of top and bottom nodes $N, M$, see Fig.~\ref{fig:s1s1_pm}{\bf d}. This result follows from the fact that the expected number of common neighbors $\overline{m}(\kappa_i, \kappa_j, \Delta\theta_{ij})$ between two nodes $i,j$ is independent of the network size~(Appendix~\ref{app:comn_nbrs_ave}). In contrast, in the degree-preserving randomized counterparts of the modeled networks, the bipartite clustering coefficient is orders of magnitude smaller, and vanishes with the network size as $\overline{c}^{\rm t,b}_{B}\sim N^{-\delta_{t,b}}$, with $\delta_{t,b} = (\gamma_{t,b}-1)/2$ (see Fig.~\ref{fig:s1s1_pm}{\bf d}), which is the expected behavior for uncorrelated bipartite networks~\cite{Kitsak2011}. A similar behavior holds for the real bipartite networks we consider (Fig.~\ref{fig:s1s1_pm}{\bf d}).

Another important property of bipartite clustering coefficient in sf/sf networks is its self-similarity with respect to a degree-thresholding renormalization procedure~\cite{Serrano2008}. Non-iterative removal of top and bottom nodes with degrees smaller than certain thresholds $(k_T, \ell_T)$ does not affect the functional form of degree-dependent bipartite clustering coefficients, which follow the same master-curve when plotted as a function of the node degree normalized by the average degree of the corresponding domain~(see Fig.~\ref{fig:all_self} and Section~\ref{app:clustering}).

Taken together, our results in this section indicate that the $\mathbb{S}^{1}\times\mathbb{S}^{1}$ model can generate a variety of bipartite network topologies, whose main characteristics are consistent with those of real bipartite systems. A natural question then is whether it is possible to reverse the synthesis, and given a bipartite network, to infer the geometric coordinates and hidden variables of its nodes, in a way congruent with the  $\mathbb{S}^{1}\times\mathbb{S}^{1}$ model. A tempting approach would be to first project the bipartite network onto one of its node domains, apply existing maximum-likelihood estimation techniques~\cite{Boguna2010, Papadopoulos2015network1, Papadopoulos2015network2} to map the resulting one-mode projection, and then use the obtained unipartite map to infer the node coordinates of the other domain~\cite{Serrano2012}. A necessary condition for this approach to work is that the geometry of the bipartite network is properly preserved in its one-mode projections. We next examine to what extend this is the case.

\section{One-Mode Projections}
\label{sec:one_mode_projections}


In one-mode projections we project a bipartite network onto one of its node domains, such that nodes of the domain are connected if they have at least one common neighbor in the bipartite network. Even though one-mode projections allow one to study bipartite networks using tools developed for unipartite networks, projections can lead to significant loss of information and artificial inflation of the projected network with fully connected subgraphs. Historically, different approaches have been proposed to deal with the loss of information. One approach, for instance, is to weigh projected links using common neighbor statistics in the original network~\cite{Battiston2004,Newman2004c,Blond2005,Morris2005}. Another approach to quantify the extent at which information is lost in one-mode projections and to identify circumstances under which one-mode projections are still acceptable is to reduce noise by identifying and removing insignificant links in the projected network~\cite{Tumminello2011,Gualdi2016,Dianati2016,Saracco2016}.

Here, we analyze the effects of one-mode projections in the context of the $\mathbb{S}^{1} \times \mathbb{S}^{1}$ model. Specifically, we ask if the latent geometry of a bipartite network is preserved in its one-mode projections. Answering this question is important, as it can shed light on how well latent geometry beneath bipartite networks can be inferred using algorithms developed for unipartite networks such as those in~\cite{Boguna2010, Papadopoulos2015network1, Papadopoulos2015network2}. In the following, we analyze the projections onto the top node domain. As before, the results for the bottom domain can be obtained by swapping the corresponding top and bottom domain variables.

The probability $r_{u}(i, j)$  that two top domain nodes $i$ and $j$ are connected in the one-mode projection, is the probability that the nodes have at least one common neighbor in the bottom domain,
\begin{equation}
r_{u}(i, j) = 1 - \prod_{1 \leq k \leq M}\left[ 1-r(\kappa_i, \theta_i; \lambda_k, \phi_k)r(\kappa_j, \theta_j; \lambda_k, \phi_k)\right],
\label{eq:uni_conn}
\end{equation}
where $k=1\ldots M$ enumerates the nodes of the bottom domain, and $r(\kappa, \theta; \lambda, \phi)$ is the connection probability in Eq.~(\ref{eq:s1_r}).

We say that the latent geometry of the bipartite network is preserved in its one-mode projection if $r_{u}(i, j)$ preserves the functional form prescribed by Eq.~(\ref{eq:r_ij}):
\begin{eqnarray}
\label{eq:proj_conn_form}
r_{u}(i,j)  &=& f\left(\frac{d_{ij}}{
d_{u}(\kappa_i, \kappa_j)}\right),\\
\nonumber d_{ij} & \propto & \Delta\theta_{ij},\\
\nonumber d_{u}(\kappa_i, \kappa_j) & \propto& \overline{k}_{u}(\kappa_{i})\overline{k}_{u}(\kappa_{j}) \propto \kappa_i \kappa_j,
\end{eqnarray}
where $f(x)$ is a monotone decreasing function of $x$, which may or may not coincide with our choice for $r(x)$ in Eq.~(\ref{eq:conn_prob}), and $d_{u}(\kappa_i, \kappa_j)$ is the characteristic distance scale for a pair of nodes with $\kappa_i$ and $\kappa_j$ in the projected network. In the case $r_{u}(i,j)$ takes the form of Eq.~(\ref{eq:proj_conn_form}) one could map projections of real bipartite networks to latent spaces using methods developed for unipartite networks. If, on the other hand, $r_{u}(i,j)$ is not in the form of Eq.~(\ref{eq:proj_conn_form}), these techniques may not map correctly bipartite networks, and they either need to be adjusted, or different techniques need to be developed.

To test if latent geometry is preserved in one--mode projections we compute $r_{u}(i,j)$ below.
We first note that since $r(\kappa_i, \theta_i; \lambda_k, \phi_k)$ and $r(\kappa_j, \theta_j; \lambda_k, \phi_k)$ depend on $\kappa_i, \theta_i$ and $\kappa_j,\theta_j$, $r_{u}(i, j)$ also depends on $\kappa_i, \theta_i, \kappa_j, \theta_j$. Due to the uniform distribution of the $\phi_k$, $r_{u}(i, j)$ does not depend on the individual values of $\theta_{i}$ and $\theta_{j}$ per se, but on the angular distance between the nodes, $\Delta\theta_{ij} = \pi - | \pi -|\theta_i - \theta_j||$. Thus, we can set, without loss of generality, $\theta_i=0$ and $\theta_j =\Delta\theta_{ij}$. Assuming a sufficiently large number of bottom nodes $M$ we can rewrite Eq.~(\ref{eq:uni_conn}) as
\begin{align}
\nonumber {\rm ln} [1- r_{u}(i, j)]  =\\
\nonumber \frac{M}{2\pi} \iint  {\rm d}\phi \, {\rm d}\lambda \, \rho(\lambda)    {\rm ln} \left[1-r(\kappa_i,0;\lambda,\phi)r(\kappa_j,\Delta\theta_{ij};\lambda,\phi)\right].\\
\label{eq:lnru_scaling}
\end{align}
Then we replace the logarithm on the right hand side of Eq.~(\ref{eq:lnru_scaling}) with its Taylor series expansion,
\begin{align}
\label{eq:uni_conn3}
\nonumber -{\rm ln} [1- r_{u}(i, j)] &=&\\
\nonumber \sum_{n=1}^{\infty} \frac{M}{2 \pi n} \int {\rm d} \lambda \, \rho(\lambda)  \int  {\rm d}\phi \, [r(\kappa_i, 0;\lambda,\phi)r(\kappa_j,\Delta\theta_{ij};\lambda,\phi)]^n.\\
\end{align}
We note that the first term of the sum in the above relation, i.e., the term corresponding to $n=1$, is the expected number of common neighbors between nodes $i$ and $j$, $\overline{m}(\kappa_i, \kappa_j, \Delta\theta_{ij})$. Second, we perform the change of integration variable $x \equiv \frac{R\phi}{\mu \sqrt{\kappa_i\kappa_j}\lambda}$, to obtain
\begin{align}
\label{eq:uni_conn5} -&{\rm ln} [1- r_{u}(i, j)] \propto  \sqrt{\kappa_i\kappa_j} \int
\lambda \, {\rm d} \lambda  \, \rho(\lambda) \sum_{n=1}^{\infty} \frac{1}{n} \times \nonumber \\
&\int_{-\infty}^{\infty} {\rm d } x  \, \left[r
\left(\sqrt{\frac{\kappa_{j}}{ \kappa_{i}}} \left|x\right| \right) r \left(\sqrt{\frac{\kappa_{i}}{ \kappa_{j}}}
\left|x-\frac{R \Delta \theta_{ij}}{ \mu \sqrt{\kappa_i\kappa_j} \lambda}\right| \right)\right]^{n}.
\end{align}

The leading contributions to the inner integrals in Eq.~(\ref{eq:uni_conn5}) come from the two maxima of each integrand at $x_1=0$ and $x_2= \Delta\widetilde{\theta}_{ij}  \equiv \frac{R \Delta\theta_{ij}}{\mu \sqrt{\kappa_i\kappa_j}\lambda}$. Specifically, for large $\Delta\widetilde{\theta}_{ij}$, connection probability $r_{u}(i,j)$ can be approximated, to the leading order, as
\begin{align}
\label{eq:uni_scaling2}
-{\rm ln} [1- r_{u}(i, j)] &\sim { \overline{\lambda^{1+\beta}} \over M^{\beta}}\left(\Delta\theta_{ij} \over d_{u}(\kappa_i, \kappa_j)   \right)^{-\beta},\\
 d_{u}(\kappa_i,\kappa_j) & = \left(\kappa_i^{\beta}\kappa_j+\kappa_i\kappa_j^{\beta}  \right) ^{1 \over \beta},\nonumber
\end{align}
where $\overline{\lambda^{1+\beta}} \equiv \int \lambda^{1+\beta}  \rho(\lambda) \, {\rm d} \lambda$ (Appendix~\ref{app:proj}). The functional form of $r_{u}(i, j)$ in Eq.~(\ref{eq:uni_scaling2}) is clearly different from that in Eq.~(\ref{eq:proj_conn_form}) since the characteristic scale $d_u(\kappa_i,\kappa_j)$ is different, indicating that latent geometry is not preserved in one--mode projections, as one could intuitively expect.

At the same time, it is important to note certain similarity between one--mode projection connection probability $r_{u}(i, j)$ and original bipartite connection probability $r(x)$. Both are decreasing functions of the angular distance $\Delta \theta_{ij}$ normalized by characteristic scales $d_{u}(\kappa_i,\kappa_j)$, albeit these scales are different in the two cases. Yet since in both cases $d_{u}(\kappa_i,\kappa_j)$ is larger for pairs of nodes with larger hidden variables, nodes characterized by larger $\kappa$ values are more likely to connect over large distances and, therefore, are expected to have larger degrees not only in the bipartite network but also in its one-mode projection, consistent with our findings in Ref.~\cite{Kitsak2011}. Furthermore, as seen from Eq.~(\ref{eq:uni_scaling2}), for sufficiently large $\frac{\Delta\theta_{ij}}{d_{u}(\kappa_i \kappa_j)} $ values $r_{u}(i,j)$  as a function of $\Delta\theta$ has the same asymptotic behavior as $r(x)$,
\begin{equation}
\label{eq:r_ij_scaling}
r_{u}(i,j) \approx -{\rm ln} [1- r_{u}(i, j)] \sim \Delta\theta_{ij}^{-\beta}.
\end{equation}

Our observation that latent geometry is not exactly preserved in one--mode projections is not specific to our choice of $r(x)$ as the connection probability function in the $\mathbb{S}^{1}\times \mathbb{S}^{1}$ model.
We show below  that the latent geometry cannot be fully preserved in one-mode projections, regardless of the functional form of $r(x)$ in Eq.~(\ref{eq:conn_prob}). Indeed, assuming that $r_{u}(i, j)$ is given by Eq.~(\ref{eq:proj_conn_form}), we can write
\begin{equation}
-{\rm ln} [1- r_{u}(i, j)] = g\left({\Delta\theta_{ij} \over
\kappa_i  \kappa_j}\right),
\end{equation}
where $g(x) \equiv -{\rm ln} [1- f(x)]$.  Next, we observe that the right hand side of Eq.~(\ref{eq:uni_conn5}) is a sum of convolutions and can be transformed into products of Fourier transforms, yielding
\begin{equation}
g(w) \propto  \int {\rm d} \lambda \, \lambda^{2} \rho(\lambda) \sum_{n=1}^{\infty} { 1 \over n } \left[r_{n}\left({\mu
\omega \lambda \over R \kappa_i} \right) r_{n} \left({ \mu \omega \lambda \over R \kappa_j } \right) \right],
\label{eq:uni_fourier}
\end{equation}
where $g(w)\equiv\int_{-\infty}^{\infty} {\rm d} x \, g(x) e^{iwx}$, and $r_{n}(w)\equiv\int_{-\infty}^{\infty} {\rm d} x \,
\left[r(\left|x\right|)\right]^{n} e^{iwx}$. Since the left hand side of Eq.~(\ref{eq:uni_fourier}) does not depend on $\kappa_i$ and $\kappa_j$ while the right hand side does, the only admissible solution is $r_{n}(w) = {\rm constant}$. This solution  corresponds to  $r(x) = \delta(x)$, where $\delta(x)$ is the Dirac delta function, and cannot be interpreted as a connection probability function.

We thus find that latent geometry cannot be fully preserved for any functional form of the connection probability function $r(x)$. At the same time, our results indicate that connection probability in one--mode projections behaves similar to that in the original network for large angular distances. This result implies that it may be possible to infer approximately the latent geometry of real bipartite networks from their one-mode projections. Yet it remains unclear how accurate such inferences can be, especially in small bipartite networks whose one--mode projections are overinflated with cliques of sizes comparable to the network size. Such problems can render geometry inference using one-mode projections highly inaccurate, especially in sf/sf networks with power law exponents $\gamma$ close to $2$, as discussed at the end of Appendix~\ref{app:proj}.

\section{Inferring latent geometry}
\label{sec:inference}

We have seen that the $\mathbb{S}^{1}\times\mathbb{S}^{1}$ model can construct synthetic bipartite networks that resemble real networks across a range of non-trivial structural characteristics, which include: (i) heterogeneity in distributions of node degrees for at least one of the two domains of nodes; (ii) power law distribution of the number of common neighbors shared between pairs of nodes; and (iii) strong bipartite clustering. These results imply that we should be able to reverse the synthesis, and given a bipartite network, to infer the hidden variables of its nodes as well as their latent distances. Below, we show that hidden variables and latent distances can be estimated from the observed node degrees and the common neighbors shared by nodes, respectively.

\begin{figure*}[!ht]
\includegraphics[width=7in]{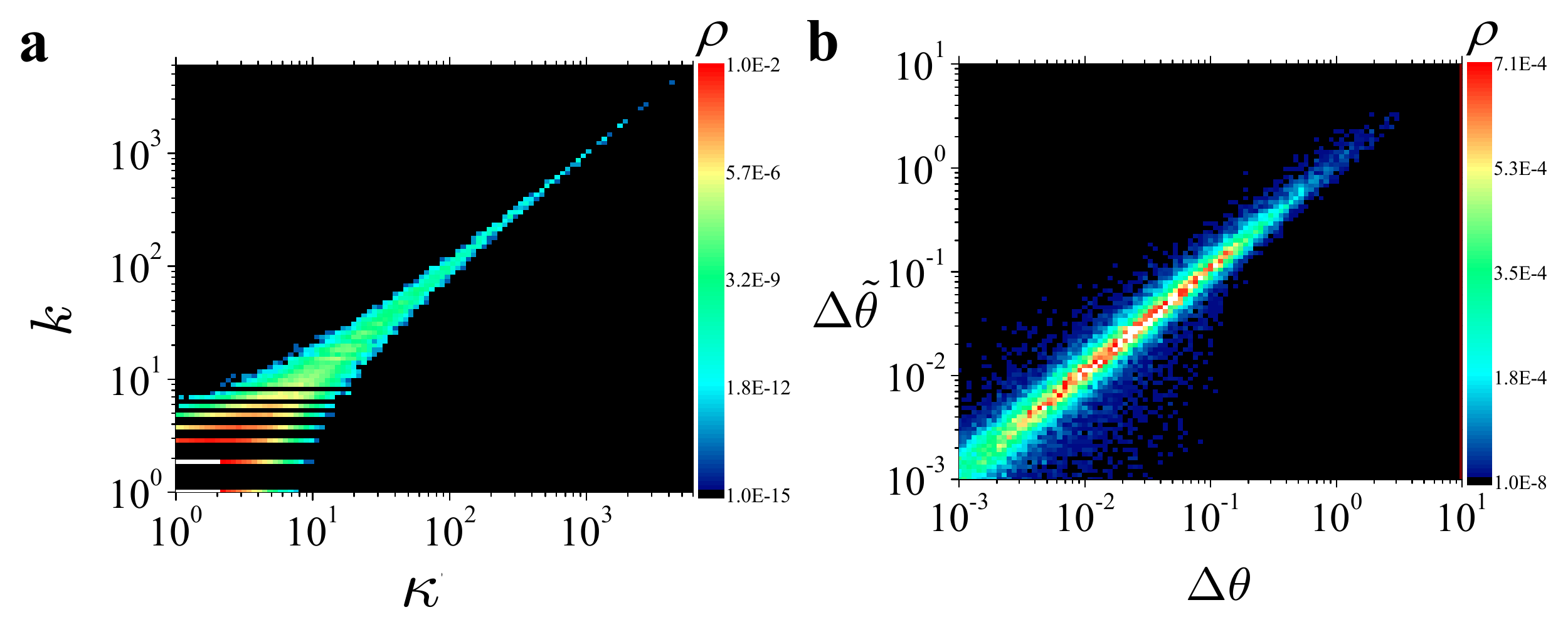}
\caption{ \footnotesize (color online) Inferring hidden variables and latent distances from the network topology. The plots correspond to an sf/ps modeled network with $N = M = 10^{5}$ nodes, $\rho(\kappa) \sim \kappa^{-2.1}$, $\overline{\kappa}=11$, $\rho(\lambda) = \delta(\lambda - 11.0)$ and $\beta = 2$. {\bf a}, The heat map displays the correlation between node degrees and hidden variables. The colors reflect the density of the nodes as shown in the legend. {\bf b}, The angular distances between the top nodes can be inferred by solving Eq.~(\ref{eq:infer_theta}). The accuracy of the inference is higher for pairs of nodes that share a larger number of common neighbors. The figure shows the inferred angular distances $\Delta\tilde{\theta}$ as a function of the true angular distances $\Delta\theta$ for pairs of top nodes with $\kappa \geq 10$ and number of common neighbors $m\geq 5$. The colors in the heat map reflect the density of the nodes as shown in the legend.
}

\label{fig:s1s1_infer}
\end{figure*}

Recall from Section~\ref{sec:basic_properties} that the resulting node degree $k_{i}$ corresponding to a particular hidden variable $\kappa_i$ is Poisson distributed with  expected value  $\overline{k}_{i}(\kappa_{i})=\kappa_{i}$. Thus, the node hidden variables can be estimated by the observed node degrees as
\begin{equation}
\kappa_{i} = \overline{k}_{i}(\kappa_{i}) \approx k_{i}.
\end{equation}
Since the variance of the Poisson distribution is equal to its mean, this estimation works better for higher degree nodes, and can be used in the case of a scale-free degree distribution in the domain, see Fig.~\ref{fig:s1s1_infer}{\bf a}.  In the case of a Poisson degree distribution in the domain, all nodes have identical hidden variables that can be estimated as
\begin{equation}
\kappa_{i} = \overline{\kappa} = \overline{k},
\label{eq:infer_k}
\end{equation}
where $\overline{k}$ is the observed average degree in the domain.

The angular distance separating two nodes can be estimated using the observed number of common neighbors between the nodes.
As shown in Section~\ref{comn_nbrs}, the number of common neighbors $m_{12}$ shared by two nodes $1,2$ is Poisson distributed with an expected value $\overline{m}_{12}\left(\kappa_1,\kappa_2, \Delta \theta_{12} \right)$ given by Eq.~(\ref{eq:paper_avg_m}), which depends on the nodes' hidden variables $\kappa_1, \kappa_2$ and their angular distance $\Delta \theta_{12}$.  If the observed number of common neighbors $m_{12}$ is sufficiently large, we can approximate $\overline{m}_{12}\left(\kappa_1,\kappa_2, \Delta \theta_{12} \right)$ as
\begin{equation}
\overline{m}_{12}\left(\kappa_1,\kappa_2, \Delta \theta_{12} \right) \approx m_{12}.
\label{eq:infer_m}
\end{equation}
The angular distance $\Delta \theta_{12}$ can be estimated using Eq.~(\ref{eq:infer_m}), which can be solved analytically for integer values of the model parameter $\beta \in (1, \infty)$ or numerically otherwise.

To test the accuracy of the proposed estimation we consider an sf/ps modeled network with $\beta = 2$. In this case, $\overline{m}_{12}$ for the top domain is given by
\begin{equation}
\label{eq:theta_estimator}
\overline{m}_{12}(\kappa_1, \kappa_2, \Delta\theta_{12}) \approx  \frac{\kappa_1 \kappa_2 \left(\kappa_1 + \kappa_2\right)} { \left( \kappa_1 + \kappa_2\right)^{2} + \left(\frac{M\Delta\theta_{12}}{2}\right)^{2}},
\end{equation}
(see Appendix~\ref{app:comn_nbrs_ave}) allowing us to estimate $\Delta \theta_{12}$ as
\begin{equation}
\Delta \theta_{12} \approx \frac{2}{M} \sqrt{\frac{k_1 k_2 \left(k_1+k_2\right)}{m_{12}} - \left(k_1 + k_2 \right)^{2}}.
\label{eq:infer_theta}
\end{equation}
We note that the above relation is an approximation and may yield angular distances outside the expected range $\Delta \theta_{12} \in [0, \pi]$ if $m_{12}$ is too large or too small.
This estimation procedure works well for pairs of nodes with large $m_{12}$ values, see Fig.~\ref{fig:s1s1_infer}{\bf b}, and it can be used for a fast estimation of the pairwise latent similarity distances between such nodes, e.g., in recommender systems.

\section{Discussion}
\label{sec:conclusion}

Understanding the organizing principles determining the structure and evolution of real bipartite networks can lead to significant advances in many challenging problems including community detection~\cite{lehmann2008biclique,wang2016asymmetric,liu2009community,benchettara2010supervised,liu2010evaluating,li2015mathematical,larremore2014efficiently,kheirkhahzadeh2016efficient}, understanding signaling pathways in gene regulatory networks~\cite{gutkind2000signaling}, multicast search~\cite{yao2015multicast}, and construction of efficient recommender systems~\cite{uchyigit2008personalization,nie2015information,an2016diffusion,pan2010weighted}.

We have shown that three common properties of many real bipartite networks---heterogeneous degree distributions, power-law distributions of the number of common neighbors, and strong bipartite clustering---appear as natural reflections of latent geometric spaces underlying these networks, where nodes are points in these spaces, while connections preferentially occur at smaller distances. The distances between nodes in the latent space can be regarded as generalized similarity measures, arising from projections of properly weighted combinations of node attributes, controlling the appearance of links between node pairs.

For our analysis, we have used the simplest possible bipartite network model with latent geometry ($\mathbb{S}^{1} \times \mathbb{S}^{1}$ model)~\cite{Serrano2012}.  Within the model, both the power law distribution of the number of common neighbors and the strong bipartite clustering emerge naturally as reflections of the metric property, i.e., the triangle inequality, of the latent space. To achieve heterogeneous degree distributions we have assigned  hidden variables to both top and bottom node domains, so that nodes with larger hidden variables connect over larger distances with higher probability and, as a result, establish more connections than nodes with smaller hidden variables.

Although not fully geometric, the $\mathbb{S}^{1} \times \mathbb{S}^{1}$ model is equivalent to the $\mathbb{H}^{2}\times\mathbb{H}^{2}$ model in Appendix~\ref{app:h2h2}, which is fully geometric, not using any hidden variables (other than node coordinates). In the $\mathbb{H}^{2}\times\mathbb{H}^{2}$ model, heterogeneous degree distributions are consequences of the exponential expansion of space in $\mathbb{H}^2$, coupled with proper boundary conditions.

As with unipartite networks, a particularly pertinent question is the possibility to infer latent geometries underlying real bipartite systems. Through the analysis of one-mode projections we have shown that latent geometry cannot be fully preserved but can be approximately preserved in one-mode projections of both sf/ps and sf/sf bipartite networks in the $\mathbb{S}^{1}\times \mathbb{S}^{1}$ model. This result supports the possibility of inferring latent geometries underlying real bipartite systems by inferring the geometries of their one-mode projections using existing techniques~\cite{Boguna2010, Papadopoulos2015network1, Papadopoulos2015network2}, as in~\cite{Serrano2012}. However, since geometry is not preserved exactly but only approximately, using one-mode projections can render geometry inference inaccurate, especially in smaller networks with weaker bipartite clustering. Such inaccuracies are particularly high in sf/sf networks with power law exponents $\gamma$ close to $2$, calling for the development of proper methods to infer latent coordinates that do not use one-mode projections. We have shown that if instead of coordinates, only pairwise latent distances between nodes with large numbers of common neighbors are to be inferred, e.g., in recommender systems~\cite{zhou2007bipartite,lu2012recommender,bobadilla2013recommender,schafer1999recommender} or in soft community detection~\cite{soft:comm,newman2015generalized}, then such inferences can be made quickly and reliably based on the common neighbor statistics.

\section{Acknowledgements}
This work was supported by NSF grants CNS-0964236 and CNS-1442999. 
F.P. also acknowledges support by the EU H2020 NOTRE project (grant 692058).
We thank M. \'{A}́. Serrano, M. Bogu\~{n}\'{a}, and P. Krapivsky for many discussions of the manuscript.

\appendix

\begin{widetext}

\section{Real-world bipartite network data}
\label{app:data}

Here we provide details on the considered real-wold bipartite networks.

a) {\it \underline{Actor-film network (IMDb)}}. The bipartite actor-film network is constructed from the Internet Movie
Database~(IMDb)~\cite{imdb_online}. The network consists of two sets of nodes: actors (top nodes) and films (bottom nodes). Every actor is connected to all films in which she/he performed. We have excluded all films whose genre is labeled by IMDb as `Adult'. The excluded `Adult' films  represent a largely isolated subset of the original actor-film network. The considered network corresponds to the period from $1960$ to $2010$.

b) {\it \underline{Condensed matter (Condmat) collaboration network}.} The Condensed matter ({\it Condmat}) collaboration network is constructed from the arXiv e-Print archive~\cite{arxiv}. The top and bottom nodes in the network are respectively authors and manuscripts published in the cond-mat section of the arXiv repository. Every manuscript is connected to all of its authors. The network includes all manuscripts published in the cond-mat section of the arXiv as of $2009$.

c) {\it \underline{High Energy Physics (HEP) collaboration network}.} The High Energy Physics ({\it HEP}) collaboration network  is also constructed from the arXiv e-Print archive~\cite{arxiv}. The top and bottom nodes in the network are respectively authors and manuscripts
published in the HEP section of the arXiv repository. Every manuscript is connected to all of its authors. The network includes all manuscripts published in the high energy physics section of the arXiv as of $2009$.

d) {\it \underline{Metabolic Network}.} The Metabolic network is based on the dataset of metabolic reactions of $107$ organisms
constructed by H. Ma and A.-P. Zeng~\cite{ma2003reconstruction}. The bipartite network consists of metabolic reactions
(bottom nodes) and metabolites (top nodes) that are connected through these reactions.

e) {\it \underline{Wikipedia}}. The constructed bipartite network of Wikipedia~\cite{wikipedia} consists of users (top nodes) and articles (bottom nodes). Users are connected to all articles they have edited. In order to eliminate random edits, we connect  users to articles which were edited by them at least three times. We consider users of the English Wikipedia who have created an account and have a user discussion page as of April $2$, $2007$. The data has been collected and processed in Ref.~\cite{crandall2008feedback}.

The basic topological properties of the considered networks are outlined in Table~\ref{si_table} and their degree distributions are shown in Fig.~\ref{fig:si_pk}.

\begin{table}[!ht]
\begin{center}
\begin{tabular}
 {|c|c|c|c|c|c|c|}
 \hline Network name& Type &$N$ & $M$ & $E$ & $\overline{k}$ & $\overline{\ell}$ \\
 \hline Actor-film network (IMDb) & sf/sf &$1602914$ & 418696 &     $6368717$ &  4.0 & 15.2 \\ 
 \hline Condensed matter (Condmat) collaboration network& sf/ps & 63799 & 79081 & 246351 & 3.9 & 3.1 \\ 
 \hline High Energy Physics (HEP) collaboration network & sf/sf & 44267 & 108907 & 255306 & 5.8 & 2.3 \\ 
 \hline Metabolic network & sf/ps & 2732 & 3568 &7750& 2.8 & 2.2 \\ 
 \hline Wikipedia & sf/sf & 45875 & 407543  & 874942 & 19.1 & 2.1\\ 
 \hline
\end{tabular}
\caption{\footnotesize Basic properties of the considered real bipartite networks. $N$ is the number of top nodes, $M$ is the number of bottom nodes, $E$ is the number of edges, $\overline{k}$ is the average top-node degree, and $\overline{\ell}$ is the average bottom-node degree.}
\end{center}
\label{si_table}
\end{table}

\begin{figure}[!ht]
\includegraphics[width=17.0 cm,angle=0]{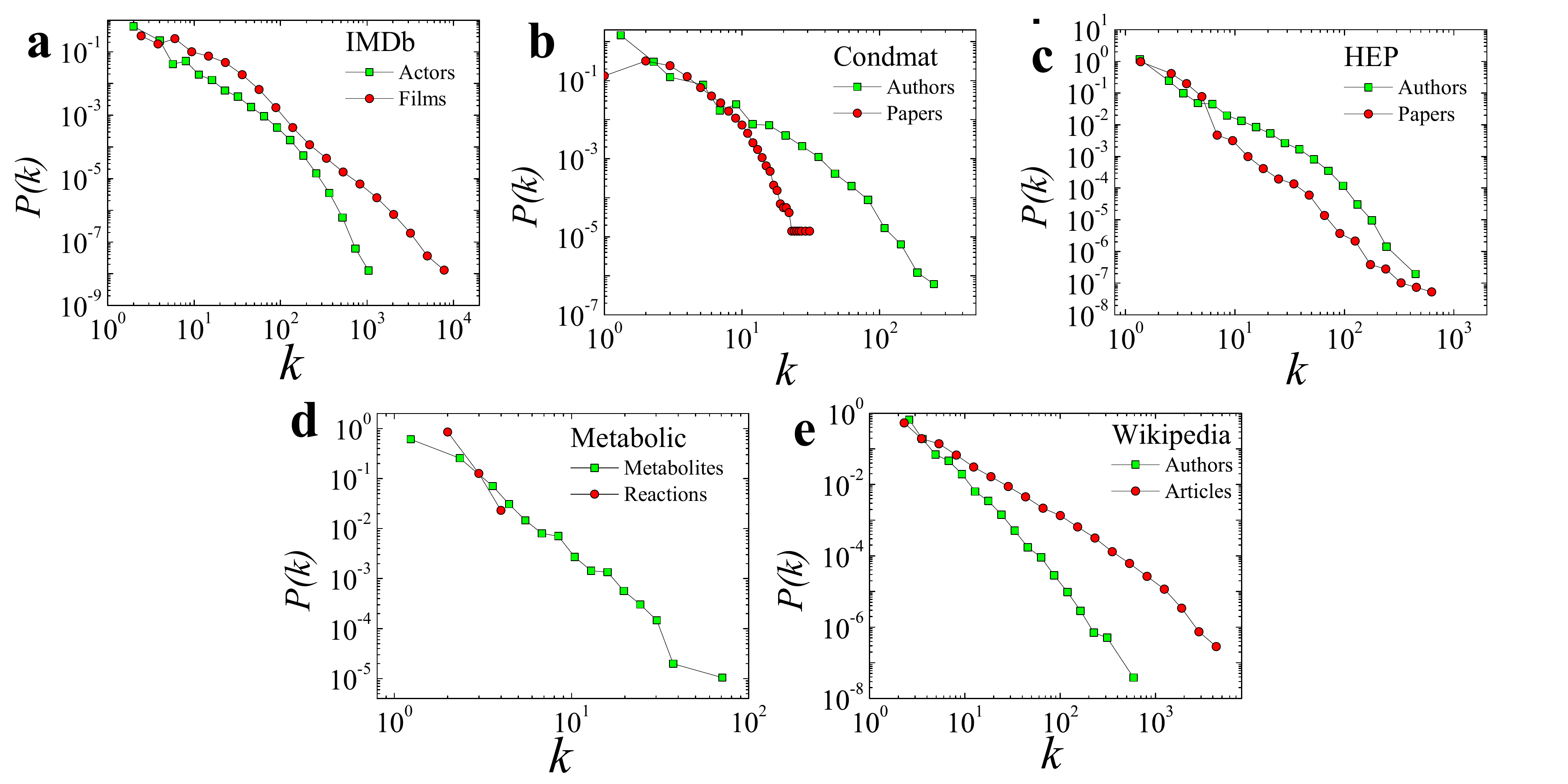}
\caption{ \footnotesize (color online) Degree distributions of top and bottom nodes of {\bf a} IMDb, {\bf b} Condmat
collaboration network, {\bf c} HEP collaboration network, {\bf d} metabolic network, and {\bf e} Wikipedia. Data have been binned
logarithmically to reduce noise.}
\label{fig:si_pk}
\end{figure}

\section{Degree-preserving randomization}
\label{app:degree_randomization}

To randomize a bipartite network while preserving the node degrees, we employ a modification of the degree-preserving randomization algorithm for unipartite networks~\cite{latapy2008basic}. Specifically, we first remove all connections in the original network. Then, we assign to every node a number of open connections equal to its degree in the original network. Finally, we add links to the network one by one, by connecting a randomly selected top-bottom node pair with open connections.  Every time a new link is established, the number of open connections of the two nodes connected by the link are decreased by one. We note that the probability to select a particular node is proportional to the number of its open connections. As a result, the randomized network is characterized by the same degree distribution as the original network and random degree-degree correlations.

\section{Bipartite networks in spaces with broken metric structure}
\label{app:non_metric_models}

The key property of any metric space is the triangle inequality: for any triplet of points $(A,B,C)$
\begin{equation}
\label{eq:triangle} d(A,C) \leq d(A,B) + d(B,C),
\end{equation}
where $d(A,B)$ is the distance between points $A$ and $B$. As explained in the main text, the triangle inequality in the underlying space leads to high clustering and power law distribution of the number of common neighbors in the observed topology of the bipartite network.

To highlight the key role of the triangle inequality, here we consider bipartite networks built in spaces with broken metric structure. To this end, we construct a bipartite network following steps $1$-$3$ of the $\mathbb{S}^{1}\times\mathbb{S}^{1}$ model (Section~\ref{sec:the_model}~A of the main text), with the difference that before connecting top-bottom nodes we randomize their distances. Specifically, we first assign node hidden variables and coordinates following steps $1, 2$ of the $\mathbb{S}^{1}\times\mathbb{S}^{1}$ model. Then, we construct the distance matrix $\{d_{ij}\} = d(\theta_{i}, \phi_{j})$ and randomize its values. To this end, we remove all distance values from the matrix and arrange them into a list of values. We then fill in the randomized matrix by drawing distance values from this list at random without replacement. Finally, we connect top-bottom nodes as in step $3$ of the  $\mathbb{S}^{1}\times\mathbb{S}^{1}$ model using the randomized distance matrix $\{d_{ij}\}$.

The above modified model preserves the values of the node hidden variables and the distribution of distances between the nodes, but breaks the triangle inequality. As seen from Figs.~\ref{fig:trg}{\bf a}, the degree distribution in bipartite networks constructed by the model with broken metric structure (non-metric) is the same as that constructed by the original $\mathbb{S}^{1}\times\mathbb{S}^{1}$ model (metric).
 At the same time, however, the  model with broken metric structure fails to reproduce the strong bipartite clustering coefficient and the fat-tail distribution of the number of common neighbors observed in the original $\mathbb{S}^{1}\times\mathbb{S}^{1}$ model~(Figs.~\ref{fig:trg}{\bf b,c}).

\begin{figure}[!ht]
\center
\includegraphics[width=5.0 cm,angle=0]{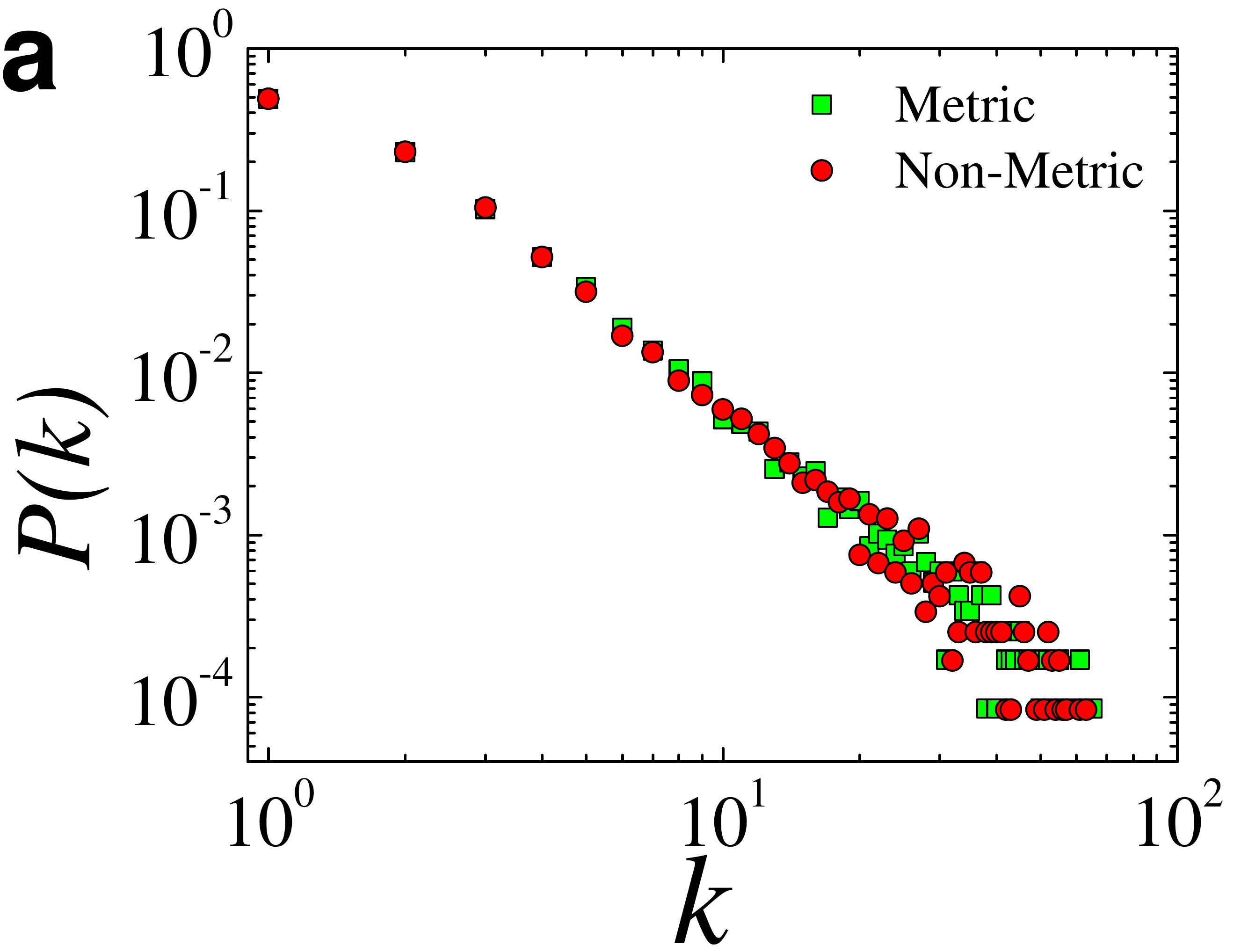}
\includegraphics[width=5.0 cm,angle=0]{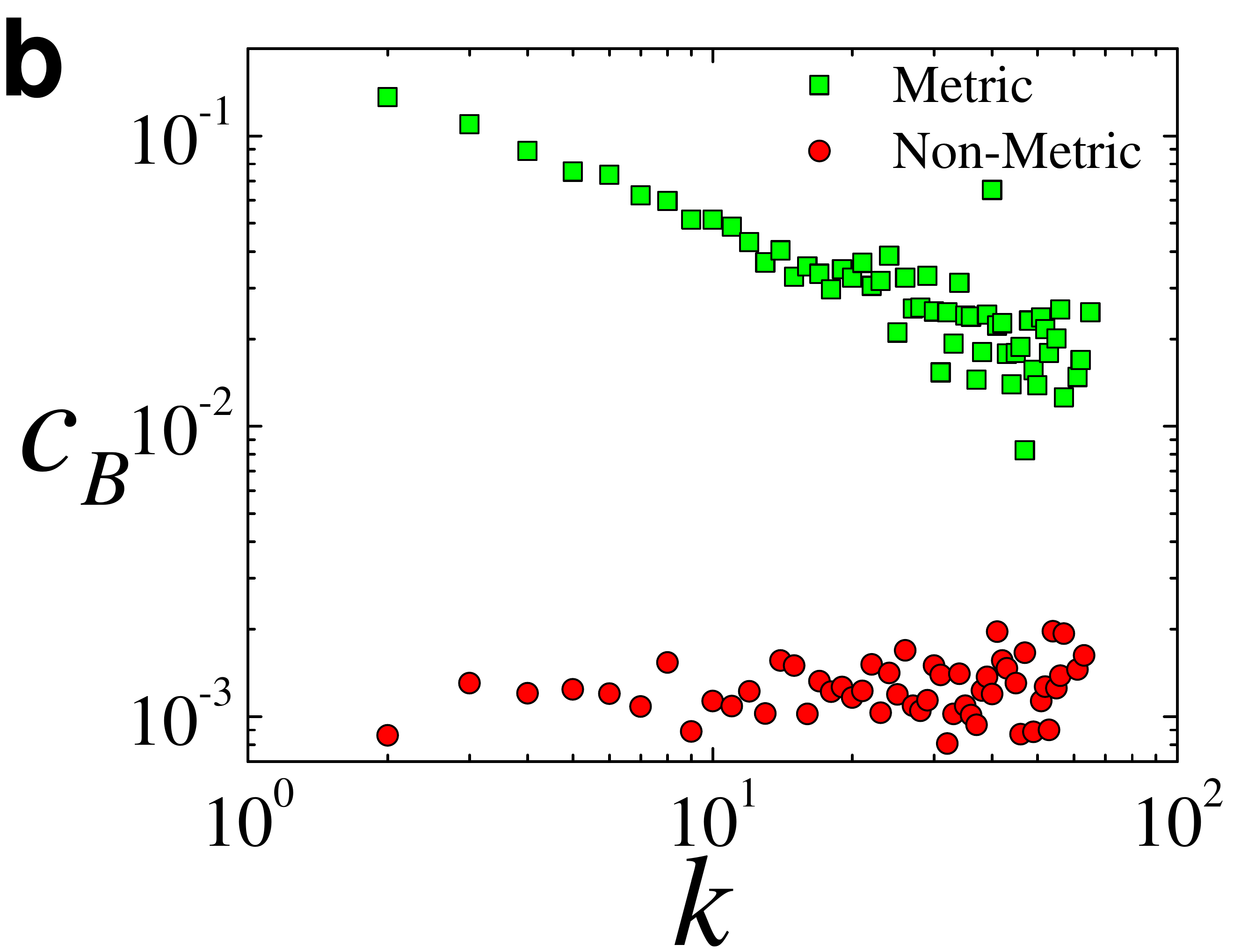}
\includegraphics[width=5.0 cm,angle=0]{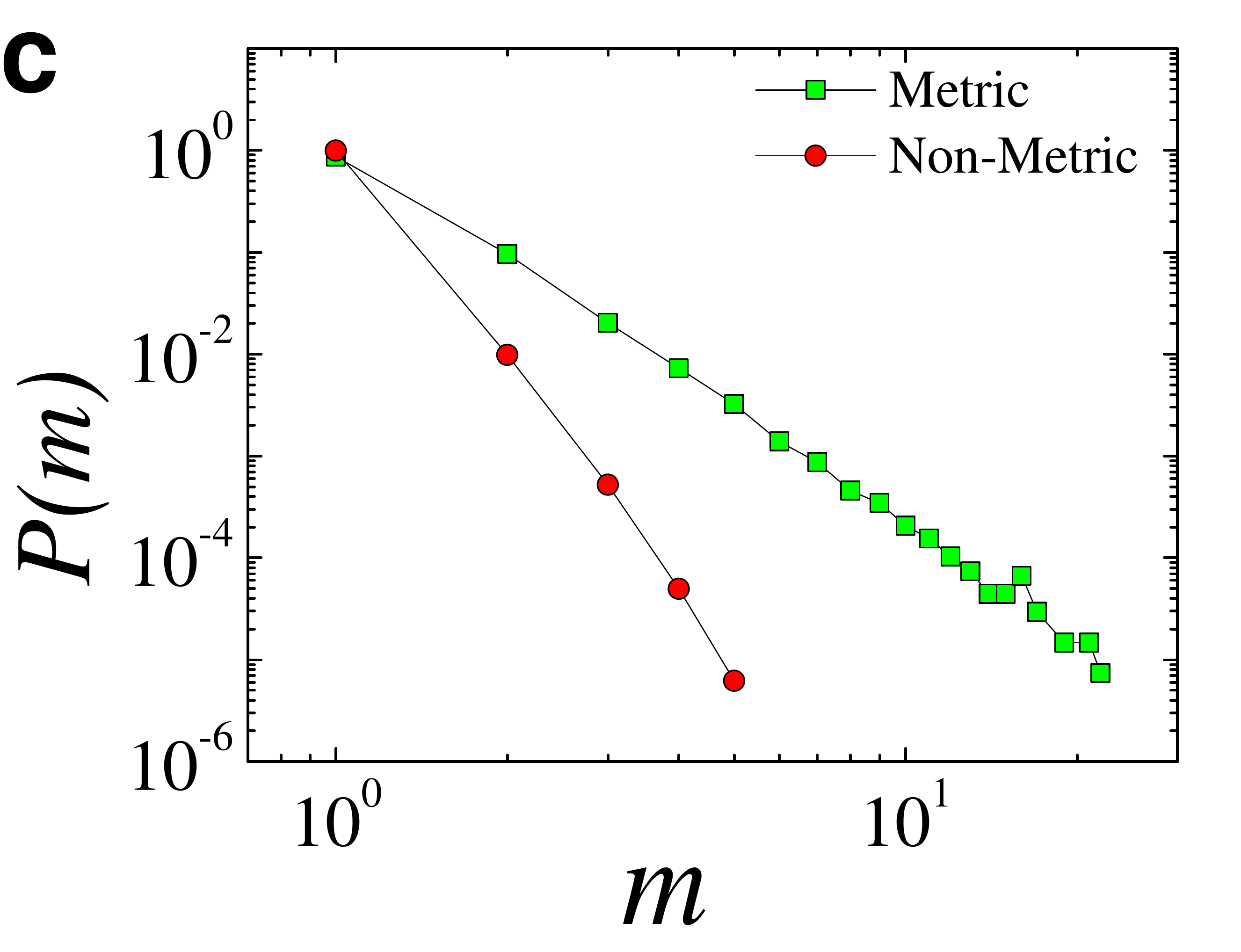}
\caption{ \footnotesize (color online) Bipartite networks in metric spaces vs. spaces with broken metric structure. {\bf a}, Degree distribution; {\bf b}, average bipartite clustering coefficient as a function of node degree; and  {\bf c}, distribution of the number of common neighbors. Both networks are sf/sf bipartite networks with $N=M=10^5$, $\gamma_1 = \gamma_2= 2.1$, $\overline{k} = \overline{\ell} = 11$  and $\beta=2.0$.
All plots correspond to the top node domain. The bottom node domain has similar properties.}
\label{fig:trg}
\end{figure}

\section{$\mathbb{H}^{2} \times \mathbb{H}^{2}$ model}
\label{app:h2h2}
In the main text we formulate the $\mathbb{S}^{1} \times \mathbb{S}^{1}$ bipartite model on the Euclidean ring $\mathbb{S}^{1}$. The choice of $\mathbb{S}^{1}$ as the latent space is made primarily for simplicity, and generalizations to other spaces are possible. Here we formulate the latent space model using a two-dimensional hyperbolic disk of constant negative curvature $K=-\zeta^{2}<0$ and radius $R_{H} \gg 1$. In the model, nodes map to points within the hyperbolic disk: angular coordinates of both node types are chosen uniformly at random, while radial coordinates are drawn from corresponding pdf functions $\rho_{t}(r)$ and $\rho_{b}(r)$. Connections between top and bottom nodes are established with probabilities depending on distances between the nodes in the hyperbolic disk. The resulting $\mathbb{H}^{2} \times \mathbb{H}^{2}$ model can be summarized as follows:
\begin{enumerate}
\item Sample the angular coordinates of top nodes $\theta_i$, $i=1,2,\ldots,N$, uniformly at random from $[0, 2\pi]$, and their radial coordinates $r_{i}$, $i=1,2,\ldots,N$, from the pdf $\rho_{t}(r)$ defined in $[0,R_H]$;
\item Sample the angular coordinates of bottom nodes $\phi_j$, $j=1,2,\ldots,M$, uniformly at random from $[0, 2\pi]$, and their radial coordinates $r_{j}$, $j=1,2,\ldots,M$, from the pdf $\rho_{b}(r)$ defined in $[0,R_H]$;
\item Connect every top-bottom node pair with probability
\begin{equation}
p(r_{i},\theta_{i};r_{j},\phi_{j}) = p\left(x_{ij}\right),\\
\end{equation}
\begin{equation}
\cosh (\zeta x_{ij}) = \cosh(\zeta r_i) \cosh(\zeta r_j) - \sinh(\zeta r_i) \sinh(\zeta r_j) {\rm cos} (\Delta \theta_{ij}).
\label{eq:app:xij}
\end{equation}
\end{enumerate}

Unlike its counterpart, the $\mathbb{H}^{2} \times \mathbb{H}^{2}$ model is purely geometric and desired degree distributions can be engineered with a proper choice of $\rho_{t,b}(r)$ and $\rho_{t,b}(\theta)$, without introducing external hidden variables. Yet, as we demonstrate below, both models are fully equivalent due to a one-to-one mapping between radial coordinates in the $\mathbb{H}^{2} \times \mathbb{H}^{2}$ model and hidden variables in the $\mathbb{S}^{1} \times \mathbb{S}^{1}$ model:
\begin{align}
r_{i} &= R_{H}-{2 \over \zeta}{\rm ln}\left(\kappa_{i} / \kappa_{0}\right),\\
r_{j} &= R_{H}-{2  \over \zeta}{\rm ln}\left(\lambda_{j}/ \lambda_{0}\right).
\label{eq:mapping}
\end{align}
It is straightforward to verify that this mapping transforms the $\mathbb{S}^{1} \times \mathbb{S}^{1}$ connection probability function into that of the $\mathbb{H}^{2} \times \mathbb{H}^{2}$ model,
\begin{eqnarray}
r\left( \frac {R\Delta\theta_{ij} }{\mu \kappa \lambda} \right) = p\left(\tilde{x}_{ij}\right),\
\label{eq:app:equiv}
\end{eqnarray}
where
\begin{equation}\label{eq:si_h2_approx_dist}
\tilde{x}_{ij}\equiv r_{i}+r_{j} + \frac{2}{\zeta} \ln (\Delta\theta_{ij}/2)
\end{equation}
is a close approximation for $x_{ij}$ in Eq.~(\ref{eq:app:xij}) for sufficiently large $\zeta r_{i}$, $\zeta r_{j}$ and $\Delta \theta_{ij} > 2\sqrt{\exp(-\zeta r_i) + \exp(-\zeta r_j)}$. Eq.~(\ref{eq:app:equiv}) establishes the equivalence between the two models since any pair of nodes $\{i,j\}$ is equally likely to be connected or disconnected in both models.

Since the two models are equivalent, the $\mathbb{H}^{2} \times \mathbb{H}^{2}$ model produces networks with the same topological properties as in the $\mathbb{S}^{1} \times \mathbb{S}^{1}$ model. In particular, sf/sf bipartite networks can be generated using $\rho_{t,b}(r) \propto \exp(\alpha_{t,b}r)$ with $\alpha_{t,b} = \frac {\zeta}{2}\left(\gamma_{t,b} - 1\right)$ and sf/ps bipartite networks can be generated using  $\rho_{t}(r) \propto \exp(\alpha_{t}r)$, $\alpha_{t} = \frac {\zeta}{2}\left(\gamma_{t} - 1\right)$, and $\rho_{b}(r)=\delta(r-R_{H})$.

First, it is important to note that the correspondence between the $\mathbb{H}^{2} \times \mathbb{H}^{2}$ and the $\mathbb{S}^{1} \times \mathbb{S}^{1}$ formulations is approximate and holds only for sufficiently large hyperbolic disk, $R_{H}\gg 1$. Indeed, when establishing the correspondence between the two models we relied on the approximation~(\ref{eq:si_h2_approx_dist}), which works well for large radial node coordinates. This is indeed the case for the majority of nodes in both sf/sf and sf/ps networks. In the case of sf/sf networks $\gamma_{t,b} > 1$, and therefore, $\alpha_{t,b} > 0$, implying that the majority of nodes is located near the periphery of the hyperbolic disk, $r_{i,j} \to R_{H}\gg 1$. The situation is the same for the top nodes of the sf/ps networks, while bottom nodes of sf/ps are all located at the edge of the hyperbolic disk, $r_{j}=R_{H} \gg 1$. Second, radial coordinates of all nodes must be positive, which is satisfied as long as  $R_{H} \geq {2 \over \zeta}{\rm ln}\left(\kappa_{\rm max} / \kappa_{0}\right)$ and  $R_{H} \geq \frac{2}{\zeta}{\rm ln}\left(\lambda_{\rm max} / \lambda_{0}\right)$, where $\kappa_{\rm max}$ and  $\lambda_{\rm max}$ are the largest hidden variable values, respectively, in the top and bottom domains of the $\mathbb{S}^{1} \times \mathbb{S}^{1}$ model. Since expected degree values in the $\mathbb{S}^{1} \times \mathbb{S}^{1}$ model are proportional to their hidden variables, Eqs.~(\ref{avg_k1},\ref{avg_l1}), $\kappa_{\rm max} \sim N$ and $\lambda_{\rm max} \sim M$, leading to
\begin{equation}
R_{H} \sim \ln (N) \sim \ln (M).
\end{equation}
%
\section{Number of common neighbors}
\label{app:comn_nbrs}

Here we derive approximate analytic expressions for the average number of common neighbors and the distribution of the number of common neighbors in the $\mathbb{S}^{1}\times\mathbb{S}^{1}$ model. To ease notation, we sometimes write the node hidden variables and angular coordinates in vector-like form: $\mathbf{x_i} = \{\kappa_i, \theta_i\}$ and $\mathbf{y_j} = \{\lambda_j, \phi_j\}$. In this case, we define the probability distribution $\rho(\mathbf{x})$ ($\rho(\mathbf{y})$) as the probability that a randomly chosen top (bottom) node is characterized by $\mathbf{x} \equiv \{\kappa, \theta\}$ ($\mathbf{y} \equiv \{\lambda, \phi\}$). Since the hidden variables are independent of the angular coordinates, $\rho(\mathbf{x}) = \rho(\kappa)\rho(\theta)$  and $\rho(\mathbf{y}) = \rho(\lambda)\rho(\phi)$. Thus, the integration (summation) over $\mathbf{x}$ or $\mathbf{y}$ corresponds to integration (summation) over both attributes: $\int {\rm d} \mathbf{x} \, \equiv \int {\rm d} \kappa \, \int {\rm d} \theta$, $\int {\rm d} \mathbf{y} \equiv \int {\rm d} \lambda \, \int {\rm d} \phi$.

\subsection{Average number of common neighbors}
\label{app:comn_nbrs_ave}

The average number of common neighbors between two nodes of the same domain can be calculated from Eq.~(\ref{eq:paper_avg_m}) in the main text. By substituting the connection probability function from Eq.~(\ref{eq:s1_r}) into Eq.~(\ref{eq:paper_avg_m}) we obtain that for large networks
\begin{equation}
\label{eq:avg_m3} \overline{m}(\kappa_1, \kappa_2, \Delta\theta_{12}) = \frac{\sqrt{\kappa_{1}\kappa_{2}}} { I \overline{\lambda}}
\int \lambda \, {\rm d} \lambda \, \rho(\lambda) \int_{-\infty}^{+\infty} {\rm d} x \, r \left(\sqrt{{\kappa_{2} \over
\kappa_{1}}} \left|x\right| \right)r \left(\sqrt{{\kappa_{1} \over \kappa_{2}}}
\left|x-\Delta \widetilde{\theta}_{12}\right| \right),
\end{equation}
where $\Delta \widetilde{\theta}_{12} \equiv N I \overline{k}\Delta \theta_{12} /( \pi \sqrt{\kappa_{1}\kappa_{2}}\lambda)$, $\Delta\theta_{12}=\pi - | \pi -|\theta_1 - \theta_2||$, and $I$ is given by Eq.~(\ref{eq:integral_beta}). It can be seen from Eq.~(\ref{eq:avg_m3}) that regardless of the functional form of $r(x)$, $\overline{m}(\kappa_1, \kappa_2, \Delta\theta_{12})$ depends on the distance between the nodes $R\Delta\theta_{12} \propto N \Delta\theta_{12}$ and not on domain size $N$, $M$ per se.

The inner integral of Eq.~(\ref{eq:avg_m3}) can be taken by residues. However, the number of poles of the integrand depends on $\beta$. A relatively simple solution exists for $\beta = 2$.  Specifically, in this case we get
\begin{equation}
\label{eq:avg_m5}
\overline{m}(\kappa_1, \kappa_2, \Delta\theta_{12}) \approx  \frac{\kappa_1 \kappa_2 \left(\kappa_1 + \kappa_2\right)} { \left( \kappa_1 + \kappa_2\right)^{2} + \left(\frac{M\Delta\theta_{12}}{2}\right)^{2}}
\end{equation}
for sf/ps bipartite networks, and
\begin{equation}
\overline{m}(\kappa_1, \kappa_2, \Delta\theta_{12}) \approx \frac{\kappa_1 \kappa_2}{\left( \kappa_1 + \kappa_2\right)}  ~_{2}F_{1}\left[1, {\gamma_b -
2 \over 2}, {\gamma_b \over 2}, - \left(\frac{N \overline{k} \Delta\theta_{12}} {2 \lambda_0 (\kappa_1 + \kappa_2)}\right)^{2}\right]
\label{second_integral}
\end{equation}
for sf/sf bipartite networks with $\rho(\lambda) = (\gamma_{b} - 1 )  \lambda_{0}^{\gamma_b - 1}\lambda^{-\gamma_{b}}$, $\lambda \in [\lambda_{0},\infty)$, where $\gamma_{b}$ is the power law degree distribution exponent of the bottom domain and $\lambda_0$ is the expected minimum bottom domain degree. The function $_{2}F_{1}$ is the Gauss hypergeometric function.

\subsection{Distribution of the number of common neighbors}
\label{app:comn_nbrs_dist}

\begin{figure}[!ht]
\includegraphics[width=6.0 cm,angle=0]{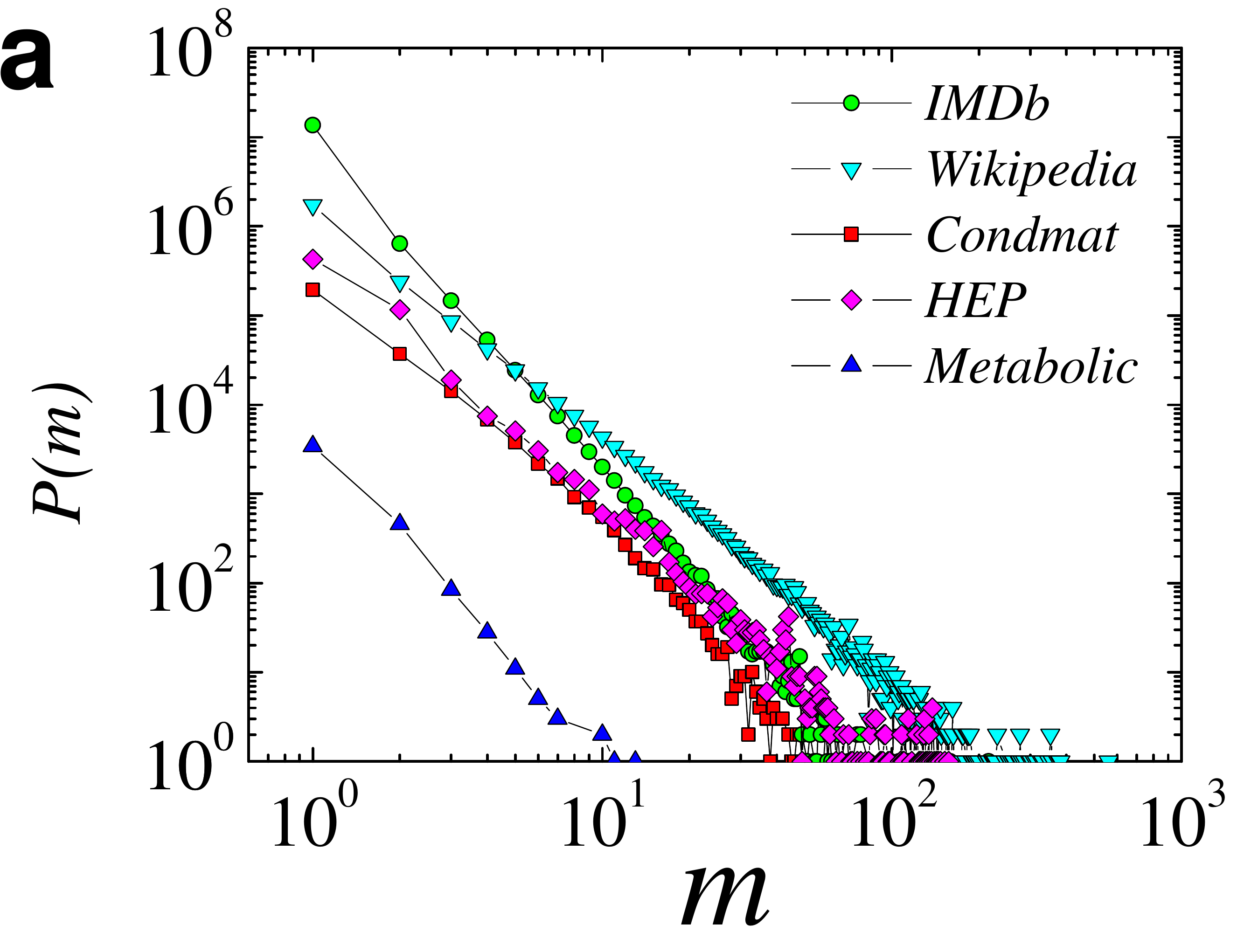}
\includegraphics[width=6.0 cm,angle=0]{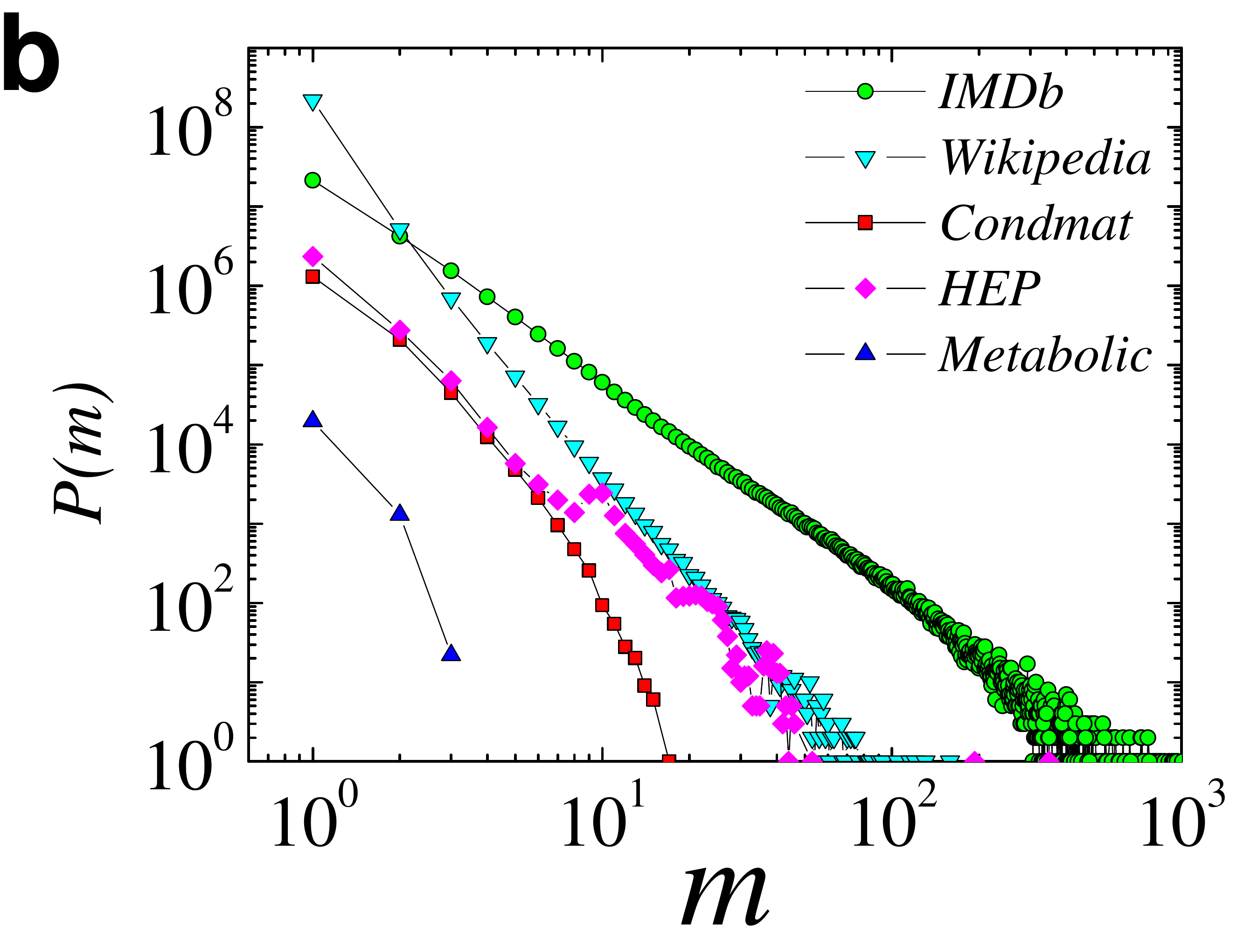}
\includegraphics[width=6.0 cm,angle=0]{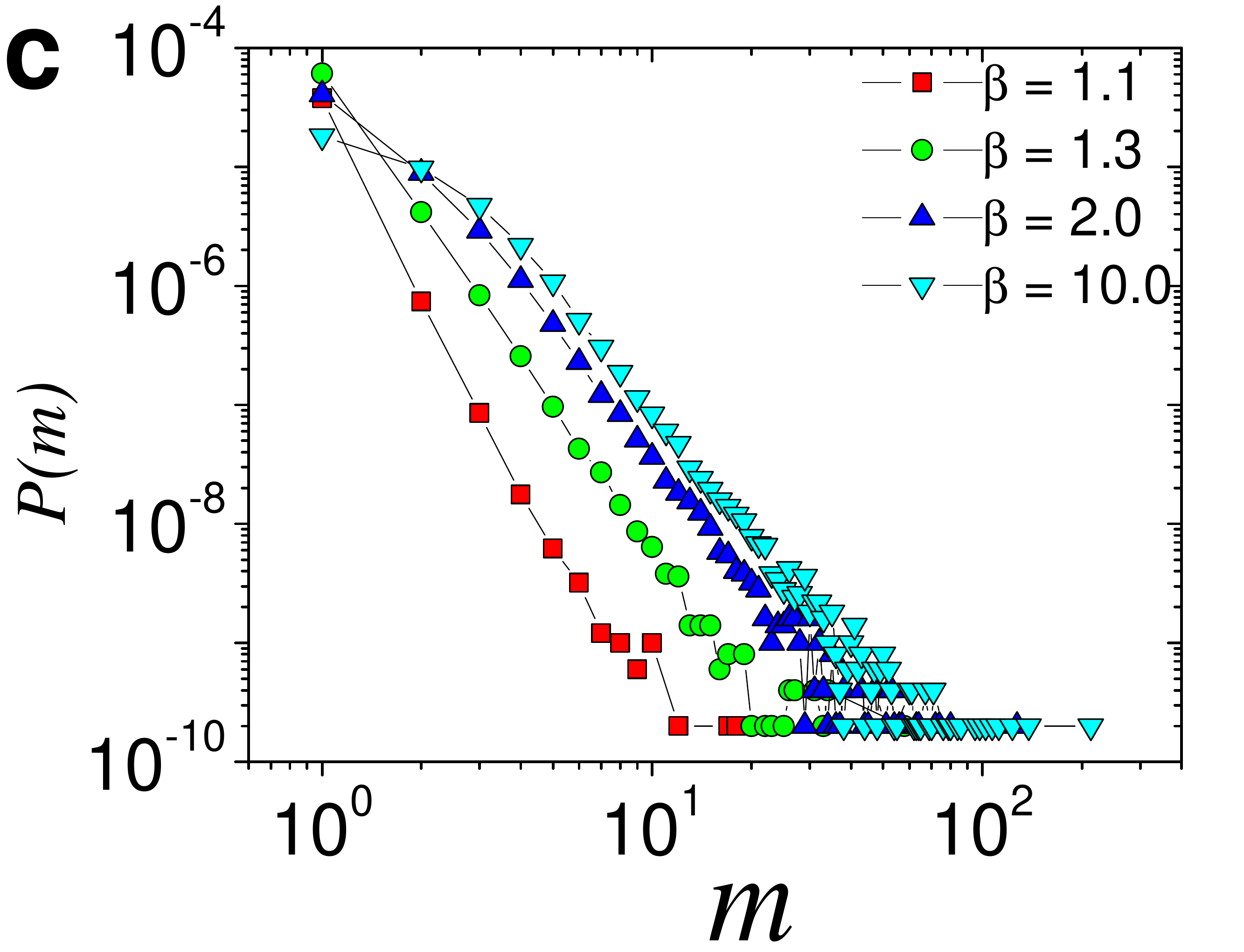}
\includegraphics[width=6.0 cm,angle=0]{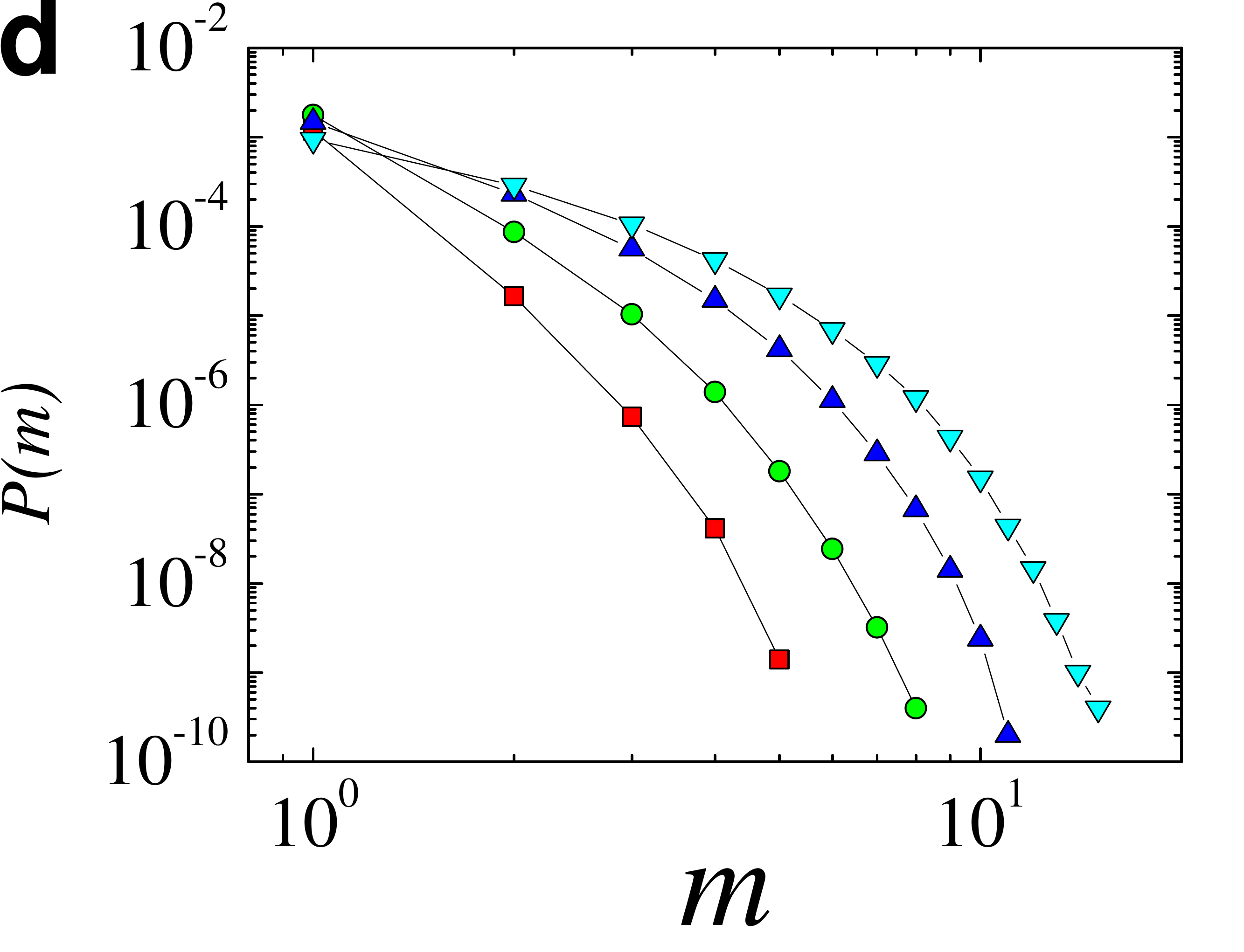}
\includegraphics[width=6.0 cm,angle=0]{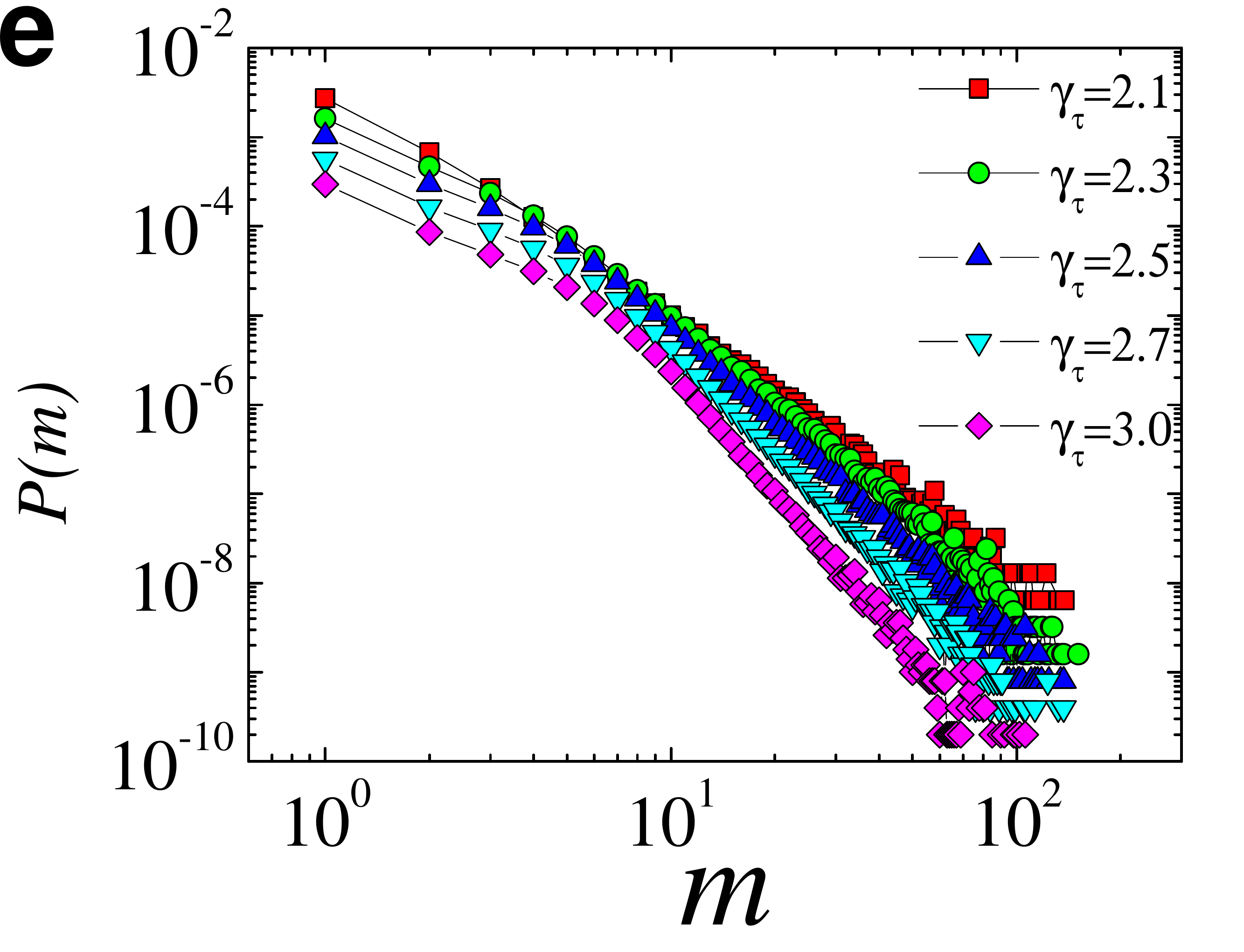}
\includegraphics[width=6.0 cm,angle=0]{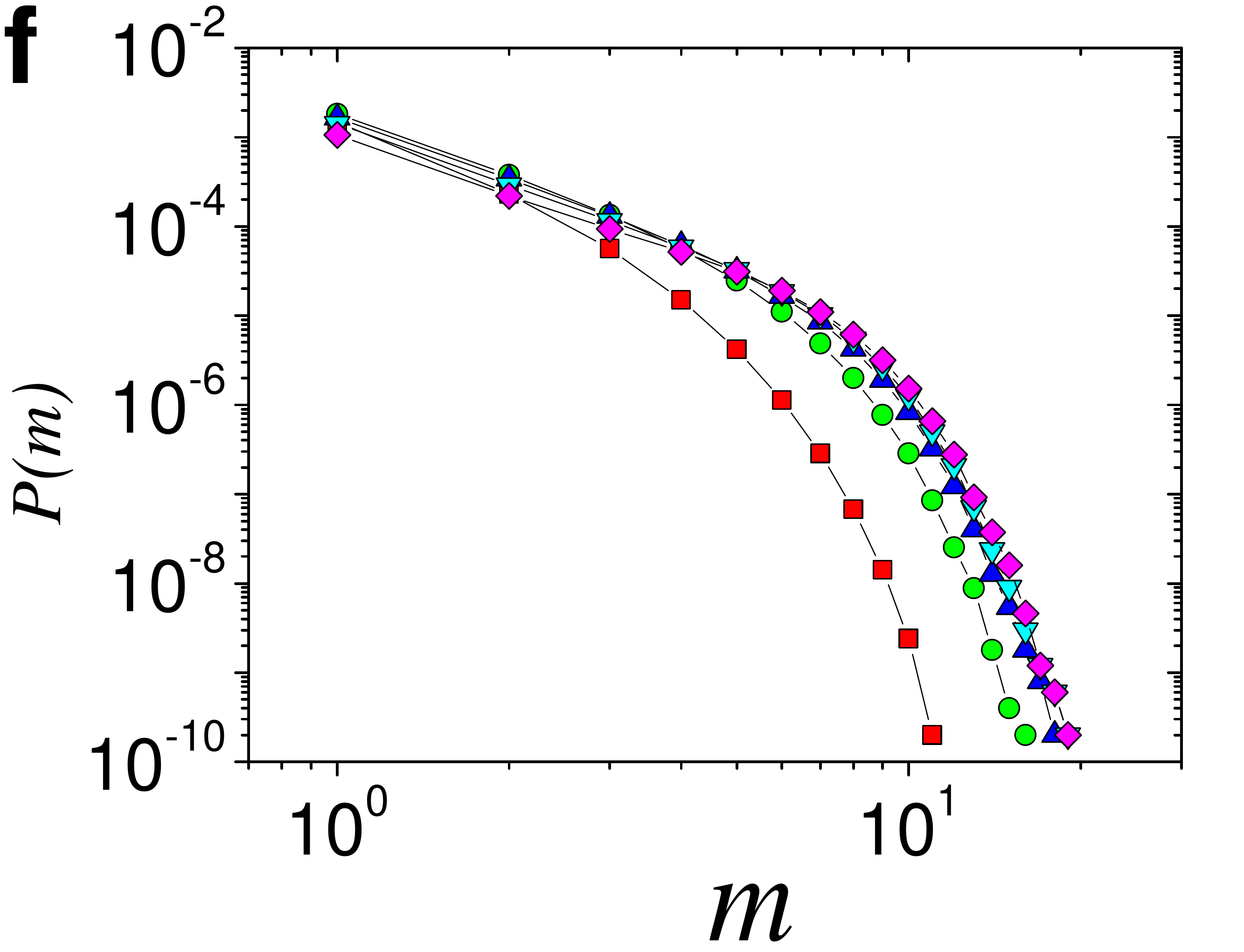}
\includegraphics[width=6.0 cm,angle=0]{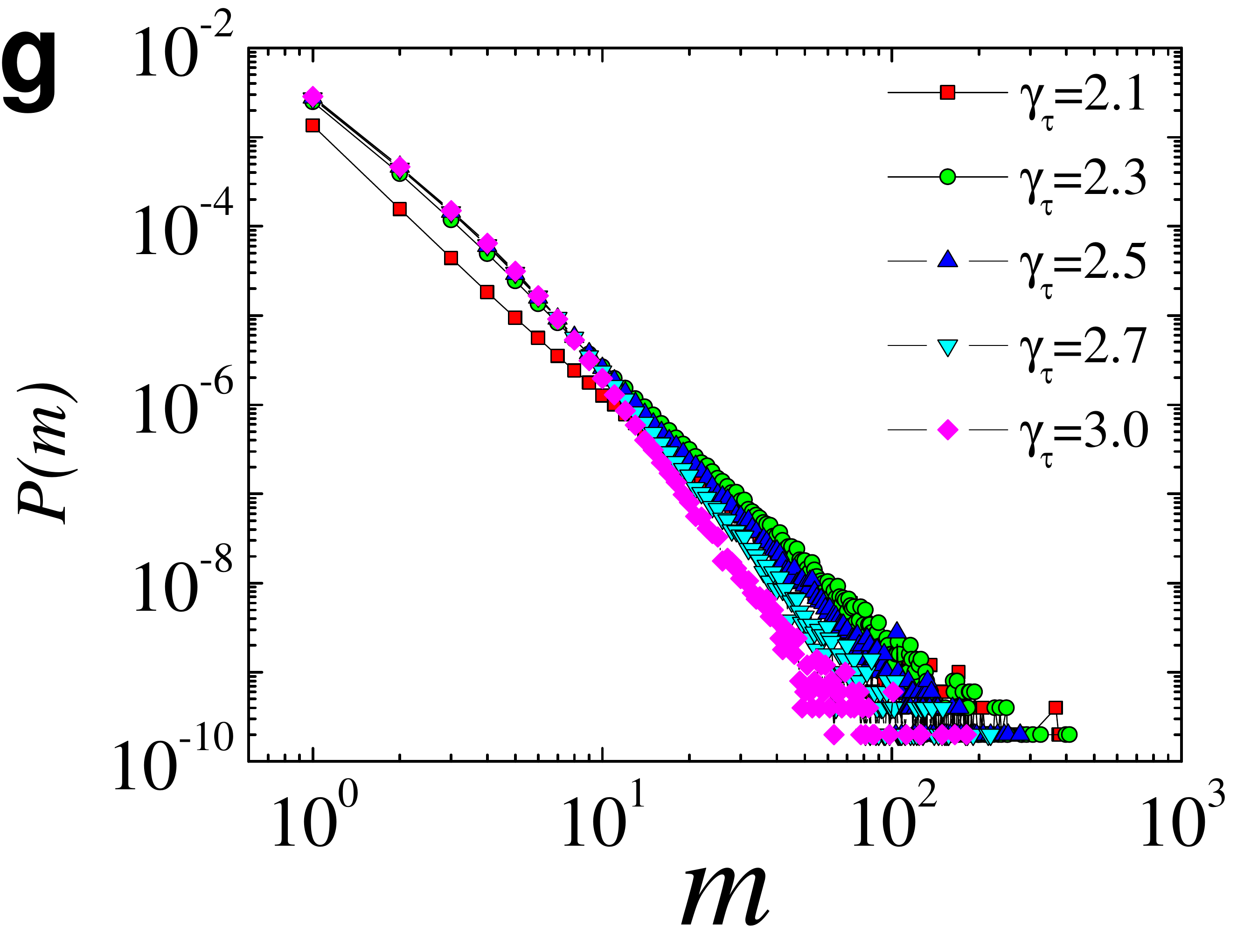}
\includegraphics[width=6.0 cm,angle=0]{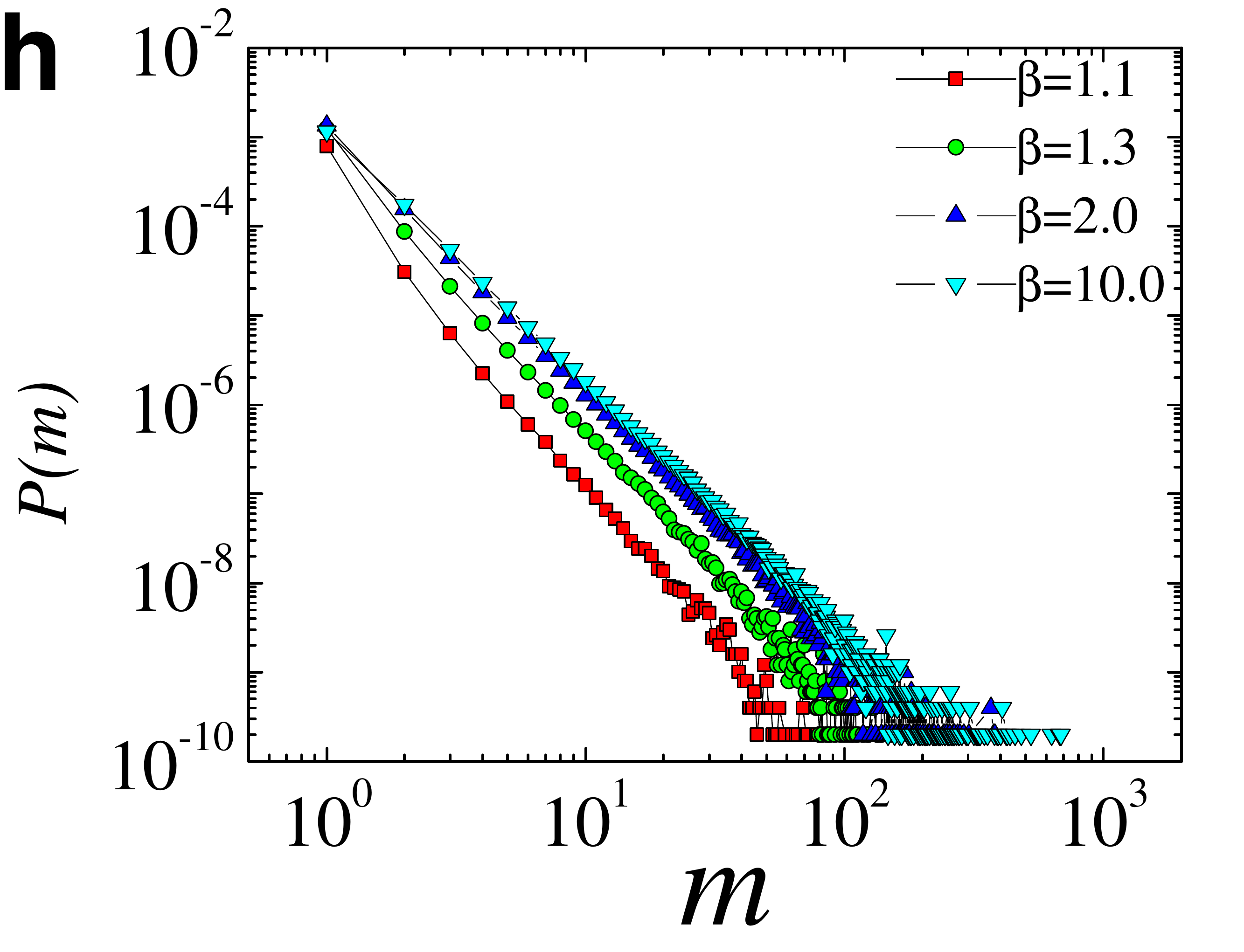}
\caption{ \footnotesize (color online) Distribution of the number of common neighbors $P(m)$ in real ({\bf a, b}) and modeled  ({\bf c-h}) bipartite networks. The properties of the real networks are outlined in Table~\ref{si_table}. Panels {\bf a, c, e, g, h} correspond to pairs of nodes in top domains, while panels {\bf b, d, f} correspond to pairs of nodes in bottom domains. The distributions $P(m)$ are not normalized for the real networks in panels {\bf a, b} (i.e., they correspond to number of nodes instead of percentage of nodes) to improve visibility. All modeled networks are characterized by $N = M= 10^{5}$. Panels {\bf c} and {\bf d} correspond to sf/ps networks with top-domain power law exponent $\gamma_{t} = 2.5$, average top-domain degree $\overline{k} = 3.0$, and different values of $\beta$. Panels {\bf e} and {\bf f} correspond to sf/ps networks with $\beta = 2.0$ and different values of $\gamma_t$, where $\overline{k}$ is chosen as $\overline{k} = {\gamma_t - 1 \over \gamma_t- 2}$. Panels {\bf g} and {\bf h} correspond to sf/sf networks. Panel {\bf g} depicts $P(m)$ for the top domain of networks that have different $\gamma_t$ values, $\overline{k}=10$, bottom-domain power law exponent $\gamma_{b}=2.5$, and $\beta=2.0$. Panel {\bf h} depicts $P(m)$ for the top domain of networks that have $\gamma_t=2.1$,  $\overline{k}=10$, $\gamma_b=2.5$, and different values of $\beta$.}
\label{fig:si_pm}
\end{figure}
We now consider the distribution of the number of common neighbors. First, we calculate the probability that two nodes with hidden variables $\kappa_1$ and $\kappa_2$ and angular coordinates $\theta_1$ and $\theta_2$ have $m$ common neighbors. To this end, we define the probability $p_{\mathbf{x_1},\mathbf{x_{2}}}(m|\mathbf{y})$ that two top nodes with vectors $\mathbf{x_1}=\{\kappa_1, \theta_1\}$ and $\mathbf{x_{2}}=\{\kappa_2, \theta_2\}$ have $m$ common neighbors among bottom nodes with vector $\mathbf{y}=\{\lambda, \phi\}$. Since pairs of nodes are connected independently with probability $r(\mathbf{x};\mathbf{y})$, $p_{\mathbf{x_1},\mathbf{x_{2}}}(m|\mathbf{y})$ is given by the binomial distribution
\begin{equation}
\label{eq:p_m_kappa} p_{\mathbf{x_1},\mathbf{x_2}}(m|\mathbf{y}) =
C^{M(\mathbf{y})}_{m}\left[r(\mathbf{x_1};\mathbf{y})r(\mathbf{x_2};\mathbf{y})
\right]^{m}\left[1-r(\mathbf{x_1};\mathbf{y})r(\mathbf{x_2};\mathbf{y}) \right]^{M(\mathbf{y})-m},
\end{equation}
where $M(\mathbf{y})$ is the total number of bottom nodes with vector $\mathbf{y}$, and $C^{k}_{l}=\left(\begin{array}{c}k\\l\\\end{array}\right)$ is the binomial coefficient. The probability that two nodes with vectors $\mathbf{x_1}$ and $\mathbf{x_2}$ have $m$ common neighbors can
be calculated as
\begin{equation}
\label{eq:pm_propagator} P_{\mathbf{x_1},\mathbf{x_2}}(m) = \sum_{\sum m_{i}=m} \prod_{i}
p_{\mathbf{x_1},\mathbf{x_2}}(m_{i}|\mathbf{y_i}),
\end{equation}
where the product goes over the entire set of the $\mathbf{y_i}$ vectors, and the summation is performed over all possible combinations of $m_{i}$ such that their sum equals $m$.

Next, we define the generating functions~\cite{wilf2013generatingfunctionology} for $p_{\mathbf{x_1},\mathbf{x_2}}(m_{i}|\mathbf{y_i})$ and
 $P_{\mathbf{x_1},\mathbf{x_2}}(m)$,
\begin{eqnarray}
\label{eq:generat}
\widehat{p}_{\mathbf{x_1},\mathbf{x_2}}(z|\mathbf{y}) \equiv \sum_{m}z^{m}p_{\mathbf{x_1},\mathbf{x_2}}(m|\mathbf{y}),\\
\widehat{P}_{\mathbf{x_1},\mathbf{x_2}}(z) \equiv
\sum_{m}z^{m}P_{\mathbf{x_1},\mathbf{x_2}}(m).
\end{eqnarray}
Since $P_{\mathbf{x_1},\mathbf{x_2}}(m)$ is given by the convolution of $p_{\mathbf{x_1},\mathbf{x_2}}(m_{i}|\mathbf{y_i})
$ (Eq.~(\ref{eq:pm_propagator})), its generating function can be expressed as the product of the generating functions of the $p_{\mathbf{x_1},\mathbf{x_2}}(m_{i}|\mathbf{y_i})$,
\begin{equation}
\label{eq:p_generat_product} \widehat{P}_{\mathbf{x_1},\mathbf{x_2}}(z) =
\prod_{i}\widehat{p}_{\mathbf{x_1},\mathbf{x_2}}(z|\mathbf{y_i}).
\end{equation}
It is easy to see that $\widehat{p}_{\mathbf{x_1},\mathbf{x_2}}(z|\mathbf{y}) =\left[1-(1-z)r(\mathbf{x_1};\mathbf{y})r(\mathbf{x_2};\mathbf{y}) \right]^{M(\mathbf{y})}$. By substituting this expression into Eq.~(\ref{eq:p_generat_product}) and taking the logarithm of both sides we get
\begin{equation}
\label{eq:p_generat2} {\rm ln} \left[\widehat{P}_{\mathbf{x_1},\mathbf{x_2}}(z) \right] = M \int {\rm d} \mathbf{y} \,
\rho(\mathbf{y}) {\rm ln} \left[(1-(1-z)r(\mathbf{x_1};\mathbf{y})r(\mathbf{x_2};\mathbf{y}) \right].
\end{equation}
One can verify that the expression for the average number of common neighbors $\overline{m}(\mathbf{x_1};\mathbf{x_2})$ (Eq.~(\ref{eq:paper_avg_m}) in the main text) can be obtained by evaluating the derivative of $\widehat{P}_{\mathbf{x_1},\mathbf{x_2}}(z)$ with respect to $z$ at $z=1$,
\begin{equation}
\label{eq:si_avg_m} \overline{m}(\mathbf{x_1};\mathbf{x_2}) = M \int {\rm d} \mathbf{y} \, \rho(\mathbf{y})
r(\mathbf{x_1}; \mathbf{y})r(\mathbf{x_2}; \mathbf{y}),
\end{equation}
as expected.
In the case of sparse bipartite networks  Eq.~(\ref{eq:p_generat2}) can be approximated as~(see Ref.~\cite{Kitsak2011})
\begin{eqnarray}
\label{eq:p_generat3}
{\rm ln} \left[\widehat{P}_{\mathbf{x_1},\mathbf{x_2}}(z) \right] \approx (z-1) \overline{m}(\mathbf{x_1};\mathbf{x_2}), \\
\label{eq:p_generat4} P_{\mathbf{x_1},\mathbf{x_2}}(m) \approx e^{-\overline{m}(\mathbf{x_1};\mathbf{x_2})} \left[
\overline{m}(\mathbf{x_1};\mathbf{x_2}) \right]^{m}/m!,
\end{eqnarray}
proving Eq.~(\ref{eq:gauss_m}) in the main text.

The distribution of the number of common neighbors $P(m)$ is obtained by averaging $P_{\mathbf{x_1},\mathbf{x_2}}(m)$ over $\mathbf{x_1}$ and $\mathbf{x_2}$,
\begin{equation}
\label{eq:Pm1} P(m) = \int \int {\rm d} \mathbf{x_1} \, {\rm d} \mathbf{x_2} \, \rho(\mathbf{x_1}) \rho(\mathbf{x_2})
P_{\mathbf{x_1},\mathbf{x_2}}(m).
\end{equation}
While in general there is no closed-form solution to Eq.~(\ref{eq:Pm1}), different closed-form solutions can be obtained for integer values of $\beta$. Below, we derive $P(m)$ in the case of sf/ps networks with $\beta = 2$. In this case, $\overline{m}(\mathbf{x_1},\mathbf{x_2})=\overline{m}(\kappa_1, \kappa_2, \Delta\theta_{12})$ is given by Eq.~(\ref{eq:avg_m5}). The first step towards computing $P(m)$ is to evaluate $P_{\kappa_1,\kappa_2}(m)$ by averaging $P_{\mathbf{x_1},\mathbf{x_2}}(m)$ in Eq.~(\ref{eq:p_generat4}) over $\Delta \theta_{12}$,
\begin{equation}
P_{\kappa_1,\kappa_2}(m) \equiv \frac{1}{\pi} \int_{0}^{\pi} {\rm d} \Delta \, \theta_{12}  P_{\mathbf{x_1},\mathbf{x_2}}(m) \propto { (\kappa_1+\kappa_2) \over\sqrt{m}m!} e^{-{\kappa_1\kappa_2 \over \kappa_1 + \kappa_2}}
\left[{\kappa_1\kappa_2 \over \kappa_1 + \kappa_2}\right]^{m}.
\end{equation}
The second step is to average $P_{\kappa_1,\kappa_2}(m)$ over all $\kappa_1$, $\kappa_2$ values. In the case of sf/ps networks $\rho(\kappa) \sim \kappa^{-{\gamma}}$, and after a series of straightforward but tedious steps we obtain
\begin{equation}
P(m) = \iint {\rm d} \kappa_1 \, \rho(\kappa_1) \, {\rm d} \kappa_2 \, \rho(\kappa_2) P_{\kappa_1,\kappa_2}(m) \propto  \frac{\Gamma[m+3-2\gamma,\kappa_0]}{m^{1/2} \Gamma[m+1] } \sim m^{3/2-2\gamma},
\label{eq:Pm_final}
\end{equation}
proving Eq.~(\ref{eq:m_power_law}) in the main text.

The analytical evaluation of $P(m)$ for arbitrary $\beta$ values is intractable due to the nontrivial dependence of  $\overline{m}(\mathbf{x_1}; \mathbf{x_2})$ on $\beta$. However, we can verify numerically that $P(m)$ is power-law distributed for different $\beta$ values, similar to real-world bipartite networks, see Fig.~\ref{fig:si_pm}.

\section{Bipartite clustering coefficient}
\label{app:clustering}

The expected value of the bipartite clustering coefficient can be derived using the hidden variable formalism for bipartite networks~\cite{Kitsak2011}. In particular, the expected bipartite clustering coefficient $\overline{c_B}(\mathbf{x})$ of a top node with vector $\mathbf{x}=\{\kappa, \theta\}$ is given by
\begin{equation}
\label{eq:c4_final}
\overline{c_B}(\mathbf{x}) = { \iint {\rm d} \mathbf{y_1} \, {\rm d} \mathbf{y_2} \,
P(\mathbf{y_1}|\mathbf{x})P(\mathbf{y_2}|\mathbf{x}) \overline{m}(\mathbf{y_1};\mathbf{y_2}) \over 2\int {\rm d}
\mathbf{y} \, P(\mathbf{y}|\mathbf{x})\overline{\ell}(\mathbf{y})- \iint {\rm d} \mathbf{y_1}\,{\rm d} \mathbf{y_2} \,
P(\mathbf{y_1}|\mathbf{x})P(\mathbf{y_2}|\mathbf{x}) \overline{m}(\mathbf{y_1};\mathbf{y_2}) },
\end{equation}
where $P(\mathbf{y}|\mathbf{x})$ is the conditional probability that a node with vector $\mathbf{x}$ is
connected to a node with vector $\mathbf{y}=\{\lambda, \phi\}$, $\overline{m}(\mathbf{y_1}; \mathbf{y_2})$ is the average number of common neighbors between nodes with vectors $\mathbf{y_1}, \mathbf{y_2}$, and $\overline{\ell}(\mathbf{y}) \equiv N\int r(\mathbf{x'}; \mathbf{y})  \rho(\mathbf{x'}) {\rm d \mathbf{x'}}$ is the expected degree of a bottom node with vector $\mathbf{y}$. The expected number of common neighbors $\overline{m}(\mathbf{y_1}; \mathbf{y_2})$ can be obtained from Eq.~(\ref{eq:paper_avg_m}) in the main text by swapping variables ($\kappa, \theta$) with ($\lambda, \phi$), while $P(\mathbf{y}|\mathbf{x})$ is given by
\begin{equation}
\label{eq:cond2}
P(\mathbf{y}|\mathbf{x}) = M {\rho(\mathbf{y}) r(\mathbf{x}; \mathbf{y}) \over \bar{k}(\mathbf{x})} = {\rho(\mathbf{y}) r(\mathbf{x}; \mathbf{y}) \over \int r(\mathbf{x}; \mathbf{y'})  \rho(\mathbf{y'}) \, {\rm d \mathbf{y'}}},
\end{equation}
where $\bar{k}(\mathbf{x}) \equiv M\int r(\mathbf{x}; \mathbf{y'})  \rho(\mathbf{y'}) \, {\rm d \mathbf{y'}}$ is the expected degree of a top node with vector $\mathbf{x}$. We note that due to the rotational symmetry of the model, $\overline{c_B}(\mathbf{x})$ depends only on the node's hidden variable $\kappa$ and not on its angular position $\theta$, that is, $\overline{c_B}(\mathbf{x}) \equiv \overline{c_B}(\kappa)$. We also note that $\overline{c_B}(\mathbf{x})$ is independent of the network size ($N \propto M$) since neither $P(\mathbf{y}|\mathbf{x})$ nor $\overline{m}(\mathbf{y_1};\mathbf{y_2})$ depend on it. Furthermore, $N$ in $\overline{\ell}(\mathbf{y})=N\int r(\mathbf{x'}; \mathbf{y})  \rho(\mathbf{x'})  \, {\rm d \mathbf{x'}}$ cancels out after performing the integration $\int r(\mathbf{x'}; \mathbf{y})  \rho(\mathbf{x'})  \, {\rm d \mathbf{x'}}$ that yields a factor $1/R \sim 1/N$, so that $\overline{\ell}(\mathbf{y})$ in the denominator of Eq.~(\ref{eq:c4_final}) does not depend on the network size either.

Finally, the average bipartite clustering of the top domain can be obtained by averaging the bipartite clustering coefficient over all nodes in the domain,
\begin{equation}
\bar{c}_B^t=\int \overline{c_B}(\mathbf{x})\rho(\mathbf{x}) \, {\rm d \mathbf{x}},
\end{equation}
and is also independent of network size. Similar results hold for the bottom node domain.

An important property of real sf/sf bipartite networks and the $\mathbb{S}^{1}\times\mathbb{S}^{1}$ model is self-similarity
of bipartite clustering coefficient with respect to a degree-thresholding renormalization procedure. Non-iterative removal of top
and bottom nodes with degrees smaller than certain thresholds $(k_T, \ell_T)$ does not affect bipartite clustering coefficient. In contrast, bipartite clustering coefficient of randomized networks increases as the threshold increases~(Fig.~\ref{fig:all_self}(A)). Moreover degree-dependent bipartite clustering coefficients preserve their functional form,
following the same master-curve when plotted as a function of the node degree normalized by the average degree of the corresponding domain:
\begin{eqnarray}
\label{eq:clust_scaling}
c_{B}^{t}(k|k_T,\ell_T) &=& f_{t}\left({k \over \overline{k}(k_T,\ell_T)}\right),\\
c_{B}^{b}(\ell|k_T,\ell_T) &=& f_{b}\left({\ell \over \overline{\ell}(k_T,\ell_T)}\right),
\end{eqnarray}
see Fig.~\ref{fig:all_self}(B).
Here $f_{t,b}(x)$ are the master curves for top and bottom domain degree-dependent clustering coefficients, while  $\overline{k}(k_T,\ell_T)$ and
$\overline{\ell}(k_T,\ell_T)$ are the average degrees of top and bottom domains after degree-thresholding. In contrast, as seen from Fig.~\ref{fig:all_self}(C), bipartite clustering coefficients of randomized versions of real and modeled bipartite networks are not self-similar.

Since expected node degrees in the $\mathbb{S}^{1}\times\mathbb{S}^{1}$ model are tuned to be equal to their hidden variables, Eqs.~(\ref{avg_k1},\ref{avg_l1}), the degree thresholding procedure is equivalent to removal of nodes with hidden variables less than threshold $(k_T, \ell_T)$. Then, the number of top and bottom nodes in thresholded network $G(k_T,\ell_T)$ decreases as
\begin{equation}
N(k_T)=N \left(\frac{k_T}{\kappa_{0}}\right)^{1-\gamma_t},~M(\ell_T)=M \left(\frac{\ell_T}{\lambda_{0}}\right)^{1-\gamma_b}, \label{eq:N_kp}
\end{equation}
where $N$ and $M$ are the number of nodes in original network $G$. Power-law distributions of hidden variables $\rho(\kappa)$ and $\rho(\lambda)$ of nodes in $G(k_T, \ell_T)$ are unchanged, but start at $k_T$ and $\ell_T$ respectively
\begin{eqnarray}
\label{eq:rho_kp}
\rho(\kappa|k_T) = (\gamma_t-1)k_T^{\gamma_t-1} \kappa^{-\gamma_t},\\
\label{eq:rho_lm} \rho(\lambda|\ell_T) = (\gamma_b-1)\ell_T^{\gamma_b-1} \lambda^{-\gamma_b}.
\end{eqnarray}
As a result, expected degrees of individual nodes and network domains scale as
\begin{eqnarray}
\label{eq:k_to_kappa} \overline{k}(\kappa|k_T,\ell_T) &=& \left(\frac{\ell_T}{\lambda_{0}}\right)^{2-\gamma_b} \kappa,\\
\label{eq:l_to_lambda} \overline{\ell}(\lambda|k_T,\ell_T) &=& \left(\frac{k_T}{\kappa_{0}}\right)^{2-\gamma_t} \lambda,\\
\label{eq:k_to_kappa_domain}\overline{k}(k_T,\ell_T) &=& \left(\frac{k_T}{\kappa_{0}}\right) \left(\frac{\ell_T}{\lambda_{0}}\right)^{2-\gamma_b}\overline{k},\\
\label{eq:l_to_lambda_domain}\overline{\ell}(k_T,\ell_T) &=& \left( \frac{\ell_T}{\lambda_{0}} \right) \left(\frac{k_T}{\kappa_{0}}\right)^{2-\gamma_t}\overline{\ell}.
\end{eqnarray}
\begin{figure}[!t]
\includegraphics[width=5.5 cm,angle=0]{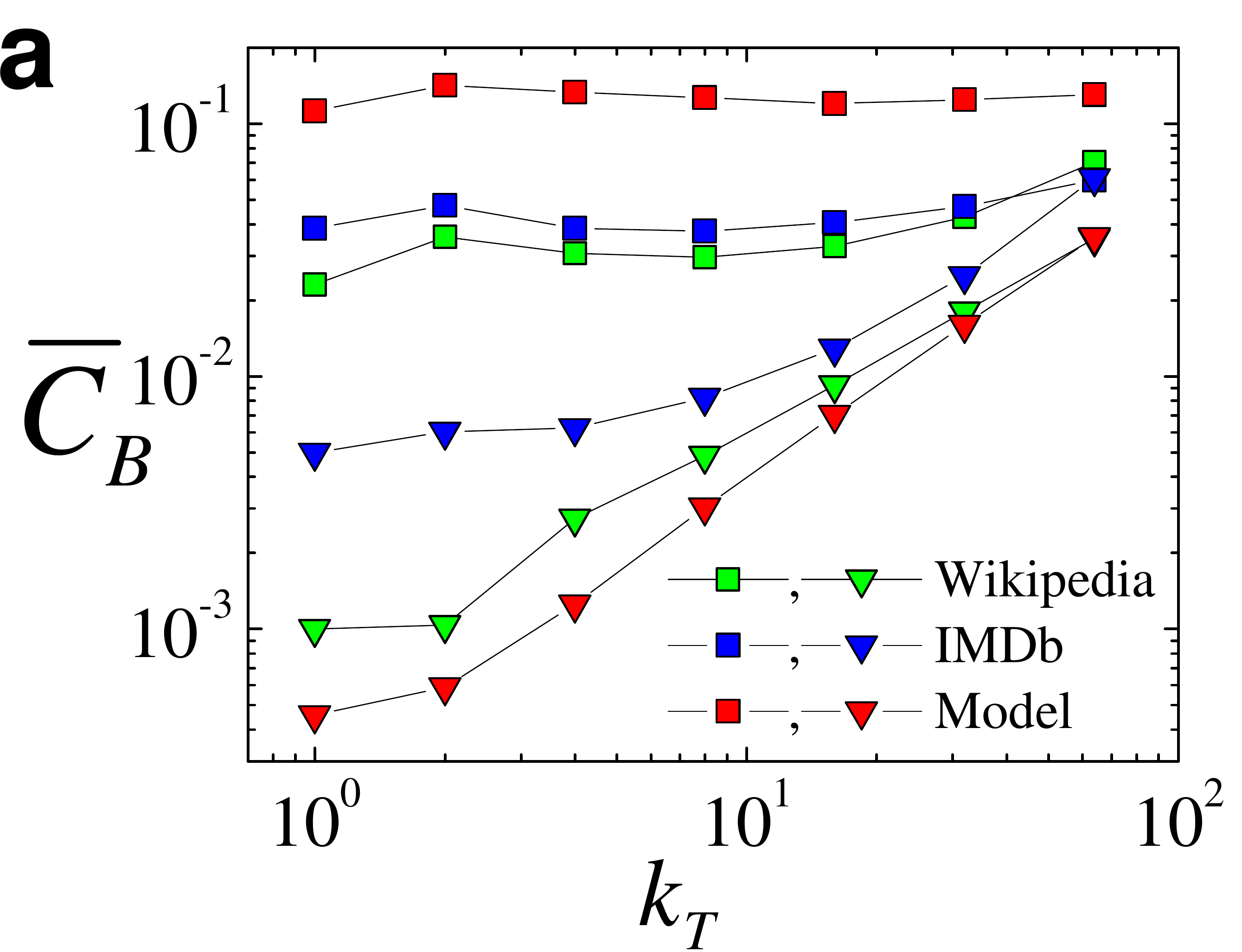}
\includegraphics[width=5.5 cm,angle=0]{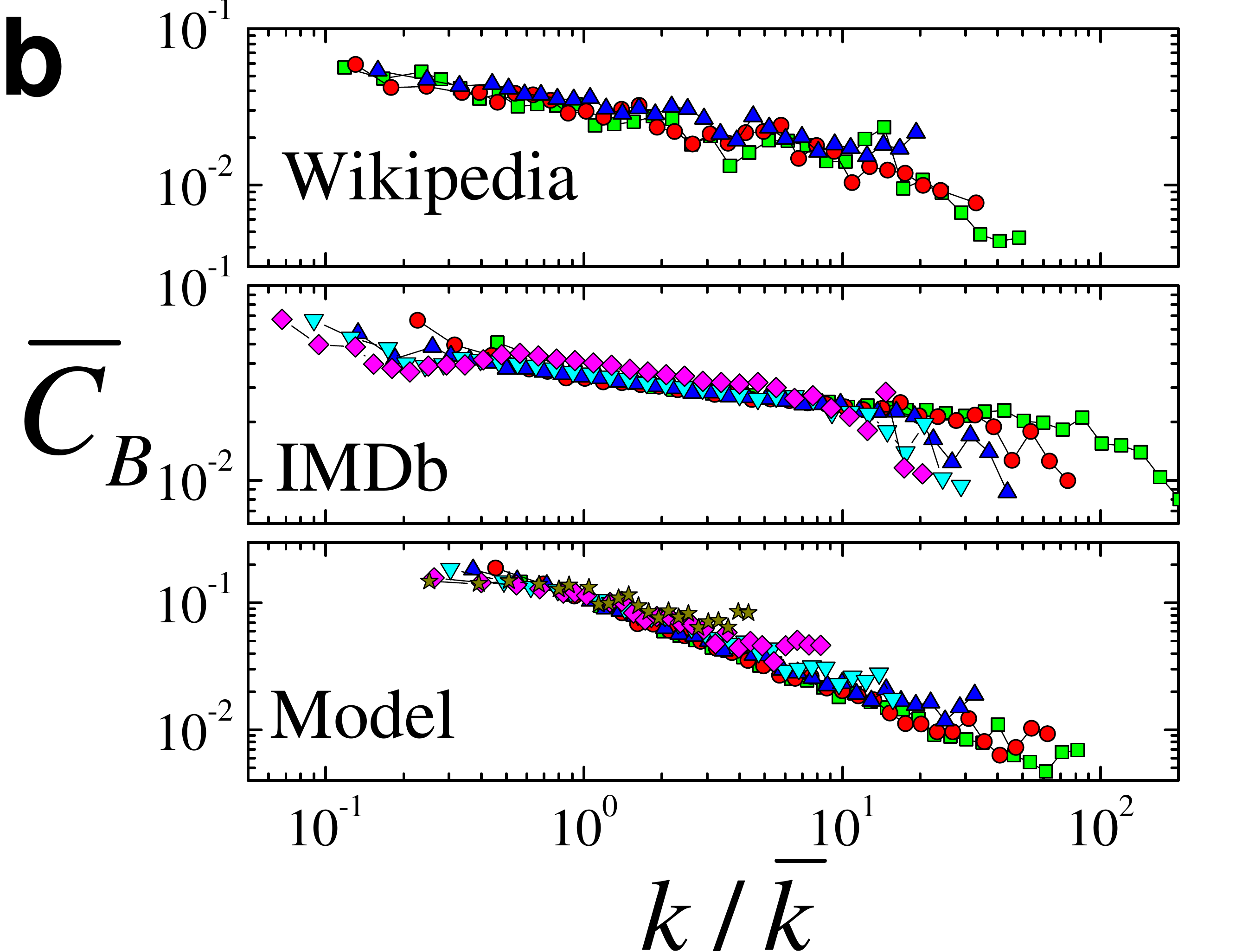}
\includegraphics[width=5.5 cm,angle=0]{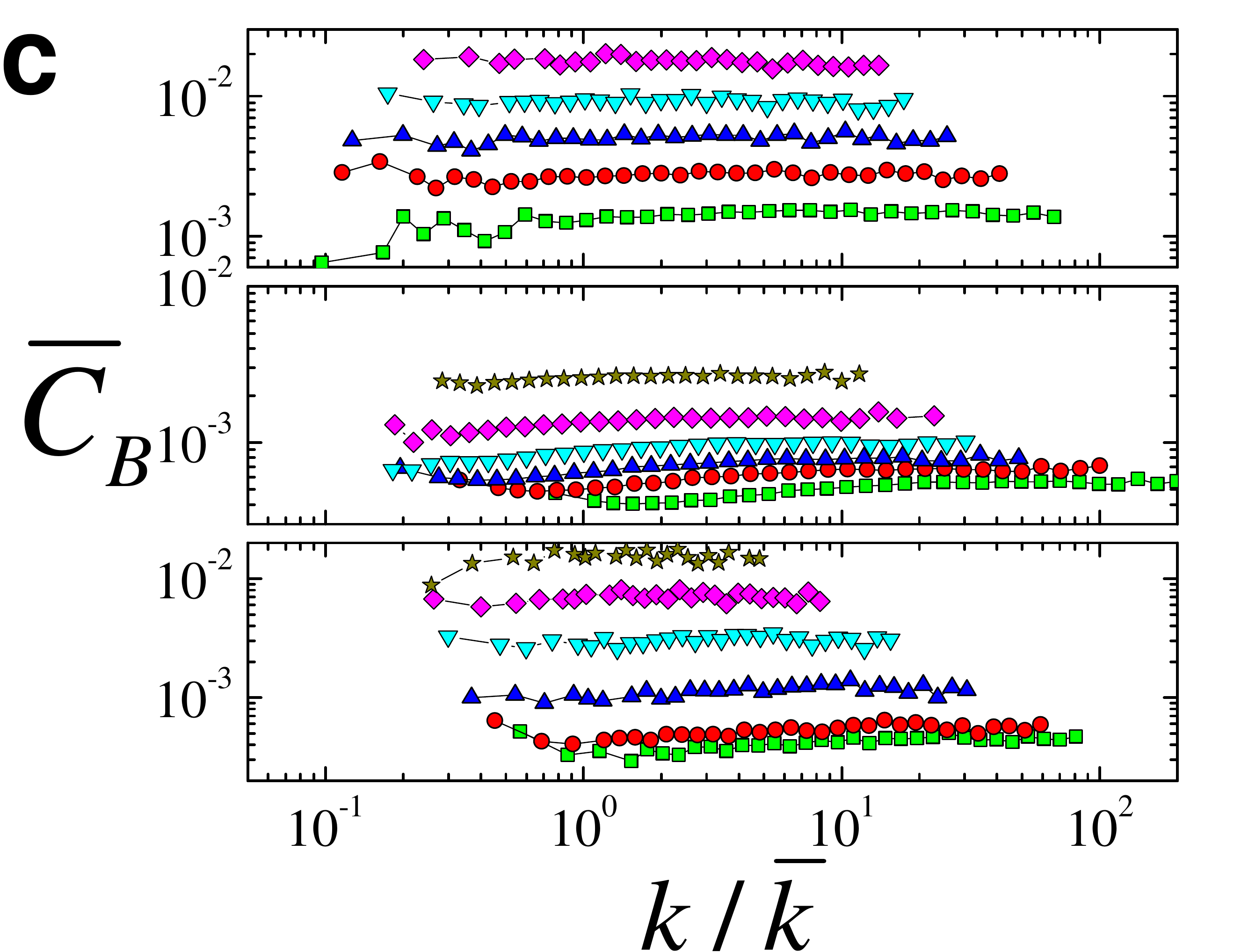}
\caption{ \footnotesize (color online) {Bipartite clustering is self-similar under degree-thresholding
renormalization. {\bf a}, Average bipartite clustering coefficient, $\overline{c}^{t}_{B}$, as a function of the threshold parameters
$k_T=\ell_T$ for the Wikipedia network (green squares), the IMDb network (blue squares), and an sf/sf modeled network
(red squares) with $N=M=10^{5}$, $\gamma_t = \gamma_b = 2.5$, $\overline{k}=\overline{\ell}=3.0$ and $\beta=2.0$. Shown with triangles are the results for
the corresponding degree-preserving randomized versions of these networks. Note that $c_{B}^{t}$ is nearly independent of
$k_T$ in the real and modeled networks, while in their degree-preserving randomized versions it increases with $k_T$.
{\bf b}, Average bipartite clustering of top nodes as a function of their normalized degree for the same real and
modeled networks as in plot~{\bf a}, calculated for different values of thresholds $k_T = \ell_T$. {\bf c}, Average
bipartite clustering of top nodes as a function of their normalized degree for the degree-preserving randomized
versions of the networks in plot~b, calculated for different threshold values $k_T = \ell_T$}.} \label{fig:all_self}
\end{figure}
Scaling relationships for degree dependent bipartite clustering coefficients follow from (\ref{eq:c4_final}). Through a number of straightforward but
tedious steps, which we omit here for brevity, we obtain

\begin{eqnarray}
\label{eq:c4_selfsimilar}
\overline{c}_{B}^{t}(\kappa|k_T,\ell_T)&=& \widetilde{f}_{t}(\kappa/k_T) = f_t\left(\frac{k}{\overline{k}(k_T,\ell_T)}\right),\\
\label{eq:c4_selfsimilar2}
\overline{c}_{B}^{b}(\lambda|k_T,\ell_T)&=& \widetilde{f}_{b}(\lambda/\ell_T) = f_b\left(\frac{\ell }{\overline{\ell}(k_T,\ell_T)}\right),
\end{eqnarray}
proving the self-similarity of sf/sf $\mathbb{S}^{1}\times \mathbb{S}^{1}$ models. Finally, it is now easy
to check, using Eqs.~(\ref{eq:c4_selfsimilar}) and (\ref{eq:c4_selfsimilar2}) that the average bipartite clustering coefficients of the top
and bottom node domains take finite values independent of the threshold $(k_T,\ell_T)$:
\begin{eqnarray}
\label{eq:avg_c41}
\overline{c}_{B}^{t}(k_T,\ell_T)= \int {\rm d} \kappa \, \rho(\kappa) \overline{c}_{B}^{t}(\kappa|k_T,\ell_T) = \overline{c}_B^{t},\\
\label{eq:avg_c42}
\overline{c}_{B}^{b}(k_T,\ell_T)= \int {\rm d} \lambda \, \rho(\lambda)\overline{c}_{B}^{b}(\kappa|k_T,\ell_T) = \overline{c}_B^{b}.
\end{eqnarray}
Unlike in sf/sf networks, bipartite clustering in sf/ps networks is self-similar only with respect to degree-thresholding of the scale-free domain, ($k_{T}$,$\lambda_{0}$).

\section{One-mode projections}
\label{app:proj}

Here we fill in the details on the estimation of the effective connection probability between nodes in one-mode projections. We consider the top node domain---similar results hold for the bottom domain. Any two nodes $i, j$ in the top domain are connected in their one-mode projection if they share at least one neighbor in the original bipartite network. Eq.~(\ref{eq:uni_conn5}) in the main text allows to estimate (up to a proportionality coefficient) the connection probability  of the two nodes in the one-mode projection, $r_{u}(i,j)$, as
\begin{equation}
 -{\rm ln} [1- r_{u}(i, j)] \propto  \sqrt{\kappa_i\kappa_j} \int
\lambda \, {\rm d} \lambda \, \rho(\lambda) \sum_{n=1}^{\infty} {1 \over n} \int_{-\infty}^{\infty} {\rm d} x \, \left[r
\left(\sqrt{{\kappa_{j} \over \kappa_{i}}} \left|x\right| \right) r \left(\sqrt{{\kappa_{i} \over \kappa_{j}}}
\left|x-{R \Delta \theta_{ij}  \over \mu \sqrt{\kappa_i\kappa_j} \lambda}\right| \right)\right]^{n}.
\label{eq:app_uni_conn5}
\end{equation}
Note that the first term in the sum of Eq.~(\ref{eq:app_uni_conn5}), i.e., the term with $n=1$, is proportional to the expected number of common neighbors between nodes $i$ and $j$. The inner integral in Eq.~(\ref{eq:app_uni_conn5}) has two maxima at $x=0$ and $x= \Delta\widetilde{\theta}_{ij} \equiv {R\Delta\theta_{ij} \over \mu \sqrt{\kappa_i\kappa_j}\lambda}$. Below, we estimate $r_{u}(i,j)$ for large $\Delta\widetilde{\theta}_{ij}$ (recall that $R \propto N \propto M$). To evaluate Eq.~(\ref{eq:app_uni_conn5}), we split the integration interval into the $5$ sub-intervals: $(-\infty, -\delta)$, $(-\delta, \delta)$, $(\delta, \Delta\widetilde{\theta}_{ij} - \delta)$, $(\Delta\widetilde{\theta}_{ij} - \delta,\Delta\widetilde{\theta}_{ij} + \delta)$, and $(\Delta\widetilde{\theta}_{ij} + \delta, \infty)$, by choosing $\delta > 0$ such that $\delta \ll \Delta\widetilde{\theta}_{ij}$ and $\delta \sim \Delta\widetilde{\theta}_{ij}$, i.e.,
\begin{equation}
\delta = c \Delta\widetilde{\theta}_{ij},
\label{eq:proj_delta}
\end{equation}
where constant $c \ll 1$.

We start by evaluating the inner integral for the $n^{th}$ term of Eq.~(\ref{eq:app_uni_conn5}) in the second sub-region. We denote this integral by $\mathcal{J}_2$. To ease notation, we drop the indices $i, j$ from the angular distance, i.e., $\Delta\theta_{ij} \equiv \Delta\theta$, $\Delta\tilde{\theta}_{ij} \equiv \Delta\tilde{\theta}$. When $\Delta\widetilde{\theta} \gg \delta$, within the integration interval $\left|x-\Delta\tilde\theta\right| \approx \Delta\tilde \theta$, and $\mathcal{J}_2$ can be approximated as
\begin{equation}
\mathcal{J}_{2}=\int
\lambda \, {\rm d} \lambda \, \rho(\lambda) \int_{-\delta}^{\delta} \, {\rm d} x \, \left[{1 \over 1 + \left( \sqrt{\kappa_j \over \kappa_i}\left| x\right|
\right)^{\beta}}\right]^{n} \left[{1 \over 1 + \left( \sqrt{\kappa_i \over \kappa_j}
\left|x-\Delta\widetilde{\theta}\right| \right) ^{\beta} }\right]^{n} \approx {2  \overline{\lambda^{1+n\beta}} \sqrt{\kappa_i \over \kappa_j}
I_{n}\left(\beta;\delta \sqrt{\kappa_j  \over \kappa_i}\right) \over\left(
\frac{R\Delta\theta}{\mu \kappa_j} \right) ^{n\beta}},
\end{equation}
where $I_{n}(\beta; a) \equiv \int_{0}^{a}  \, {{\rm d} x  \, \over \left(1 + x^{\beta}\right)^{n} }$ and $\overline{\lambda^{1+n\beta}} \equiv \int \lambda^{1+n\beta} \rho(\lambda) \, {\rm d} \lambda$. $I_{n}(\beta; a)$ increases as a function of $a$ and it is bounded from above as
\begin{equation}
I_{n}(\beta; a) \leq I_{n}(\beta; \infty) = {\Gamma\left[n - 1/ \beta \right]\Gamma\left[1+1/\beta\right] \over
\Gamma\left[ n \right]},
\label{eq:integral2}
\end{equation}
where $\Gamma[x]$ is the gamma function. Depending on $\rho(\lambda)$, $\overline{\lambda^{1+n\beta}}$ may depend on $M$. Since $R \propto M$ and the other terms in Eq.~(\ref{eq:integral2}) are independent of the network size, we can write
\begin{equation}
\mathcal{J}_{2} \sim \overline{\lambda^{1+n\beta}} M^{-n \beta}.
\label{j_2_scaling}
\end{equation}
The evaluation of the integral in the fourth sub-region, $\mathcal{J}_4$, is the same as $\mathcal{J}_{2}$ but with the $\kappa_i$ and $\kappa_j$ variables swapped,
\begin{equation}
\mathcal{J}_{4}=\int
\lambda  \rho(\lambda) \, {\rm d} \lambda \, \int_{\Delta\widetilde{\theta}-\delta}^{\Delta\widetilde{\theta}+\delta} {\rm d} x \, \left[{1 \over 1 + \left(
\sqrt{\kappa_j \over \kappa_i} \left|x\right| \right)^{\beta}}\right]^{n} \left[{1 \over 1 + \left( \sqrt{\kappa_i
\over \kappa_j} \left|x-\Delta\widetilde{\theta}\right| \right) ^{\beta} }\right]^{n} \approx {2 \overline{\lambda^{1+n\beta}} \sqrt{\kappa_j \over
\kappa_i} I_{n}\left(\beta;\delta \sqrt{\kappa_i \over \kappa_j}\right) \over\left(
\frac{R\Delta\theta}{\mu \kappa_i} \right) ^{n\beta}}.
\label{j_4_scaling}
\end{equation}
Therefore, similar to $\mathcal{J}_{2}$,
\begin{equation}
\mathcal{J}_{4} \sim \overline{\lambda^{1+n\beta}} M^{-n \beta}.
\label{j_4_scaling2}
\end{equation}

The integrals in the fifth and third sub-regions, $\mathcal{J}_5$, $\mathcal{J}_3$, are evaluated below
\begin{align}
\mathcal{J}_{5} &= \int
\nonumber \lambda  \rho(\lambda) \, {\rm d} \lambda \,\int_{\Delta\widetilde{\theta}+\delta}^{\infty} {\rm d} x \left[{1 \over 1 + \left(\sqrt{\kappa_j \over
\kappa_i} \left|x\right| \right)^{\beta}}\right]^{n} \left[{1 \over 1 + \left(\sqrt{\kappa_i \over \kappa_j}
\left|x-\Delta\widetilde{\theta}\right| \right) ^{\beta} }\right]^{n} \\
&\propto \int \lambda \rho(\lambda) \, {\rm d} \lambda
\int_{\Delta\widetilde{\theta}+\delta}^{\infty} \, {{\rm d} \, x \over x ^{n\beta} ( x-\Delta\widetilde{\theta})^{n\beta} }
\leq  \int \lambda  \rho(\lambda) \, {\rm d} \lambda  \int_{\Delta\widetilde{\theta}+\delta}^{\infty}  {{\rm d} x  \over x ^{n\beta} \delta ^{n\beta} }
 \leq \int \lambda  \rho(\lambda) \, {\rm d} \lambda   \int_{\Delta\widetilde{\theta}}^{\infty} {{\rm d} x \over x ^{n\beta} \delta ^{n\beta} }
\sim \overline{\lambda^{2n\beta}}  M^{1-2n\beta},
\label{j_5_scaling}
\end{align}
where the last scaling holds due to  Eq.~(\ref{eq:proj_delta}), and
\begin{equation}
\mathcal{J}_{3} \propto \int
\lambda \rho(\lambda)\,{\rm d} \lambda  \int_{\delta}^{\Delta\widetilde{\theta}-\delta}{{\rm d} x \over x ^{n\beta} (\Delta\widetilde{\theta}-x)
^{n\beta} } \leq  \int
\lambda  \rho(\lambda) \, {\rm d} \lambda \, \int_{\delta}^{\Delta\widetilde{\theta}-\delta}{{\rm d} x \over \delta ^{2n\beta} } \sim \overline{\lambda^{2 n\beta}}  M^{1-2n\beta}.
\label{j_3_scaling}
\end{equation}
The integral in the first sub-region, $\mathcal{J}_{1}$ can be approximated similar to $\mathcal{J}_{5}$, resulting in
\begin{equation}
\mathcal{J}_{1}  \sim \overline{\lambda^{2n\beta}}  M^{1-2n\beta}.
\label{j_1_scaling}
\end{equation}

In the case of sf/ps networks $\rho(\lambda) = \delta(\lambda - \overline{\lambda})$ and therefore, all moments of $\rho(\lambda)$ are independent of $M$. In the case of sf/sf networks $\rho(\lambda) \sim \lambda^{-\gamma}$, $\gamma > 2$, and sufficiently high moments of $\rho(\lambda)$ are size dependent. In a finite sample of randomly drawn $\lambda$ values, the maximum value $\lambda_{\rm max}$ can be estimated with the natural cut off, $\lambda_{\rm max} \sim M^{\frac{1}{\gamma-1}}$~\cite{dorogovtsev2002evolution}. The natural cut-off allows one to estimate size-dependent moments of  $\rho(\lambda)$ as
\begin{eqnarray}
\label{eq:app:moment1}
\overline{\lambda^{1+n\beta}} \propto \lambda_{\rm max}^{2+n \beta - \gamma} &\sim& M^{\frac{2+n \beta - \gamma}{\gamma-1}},\\
\overline{\lambda^{2n \beta}} \propto \lambda_{\rm max}^{1+2 n \beta - \gamma} &\sim& M^{\frac{1+2n \beta - \gamma}{\gamma-1}}.
\end{eqnarray}

It is straightforward to verify using Eqs.~(\ref{j_2_scaling}),(\ref{j_4_scaling}),(\ref{j_5_scaling}),(\ref{j_3_scaling}), and (\ref{j_1_scaling}) that $\mathcal{J}_{1}, \mathcal{J}_{3}$ and $\mathcal{J}_{5}$ decrease faster than $\mathcal{J}_{2}$ and $\mathcal{J}_4$ as $M$ increases, for both sf/ps and sf/sf bipartite networks. Therefore, to the leading order, $r_{u}(i,j)$  can be approximated as
\begin{equation}
-{\rm ln} [1- r_{u}(i,j)] \propto  \frac{\overline{\lambda^{1+\beta}}} { \left(M \Delta \theta_{ij}\right)^{\beta}}  \left[\kappa_i \left(\kappa_j\right)^{\beta}  I_{1}\left(\beta;
\delta\sqrt{{\kappa_j\over \kappa_i}}\right)   + \kappa_j \left(\kappa_i\right)^{\beta}I_{1}\left(\beta; \delta\sqrt{{\kappa_i\over \kappa_j}}\right)
\right],
\label{eq:app:ru}
\end{equation}
where $\overline{\lambda^{1+\beta}} \equiv \int  \lambda^{1+\beta} \rho(\lambda) \, {\rm d} \lambda$. It is important to note that since this leading order corresponds to the $n=1$ term in  Eq.~(\ref{eq:app_uni_conn5}), and since this term is proportional to the expected number of common neighbors between the two nodes,
\begin{equation}
-{\rm ln} [1- r_{u}(i,j)] \propto \overline{m}(\kappa_i, \kappa_j, \Delta\theta_{ij}),
\end{equation}
then the right hand side of Eq.~(\ref{eq:app:ru}) is proportional to the expected number of common neighbors between nodes $i$ and $j$, proving Eq.~(\ref{eq:avg_m_vs_dtheta}) in the main text.

Finally, since $\delta \sim \Delta \tilde{\theta} \gg 1$, $I_{1}\left(\beta; \delta\sqrt{{\kappa_j\over \kappa_i}}\right) \approx I_{1}\left(\beta; \delta\sqrt{{\kappa_i\over \kappa_j}}\right) \approx I_{1}(\beta;\infty)$, and therefore
\begin{align}
\label{eq:lnru}
-{\rm ln} [1- r_{u}(i, j)] &\sim { \overline{\lambda^{1+\beta}} \over M^{\beta}  }\left(\Delta\theta_{ij} \over d_{u}(\kappa_i, \kappa_j)   \right)^{-\beta},\\
 d_{u}(\kappa_i,\kappa_j) & \equiv \left(\kappa_i^{\beta}\kappa_j+\kappa_i\kappa_j^{\beta}  \right) ^{1 \over \beta}\nonumber,
\end{align}
proving Eq.~(\ref{eq:uni_scaling2}) in the main text.

Equation~(\ref{eq:lnru}) allows to analyze the behavior of $-{\rm ln} [1- r_{u}(i,j)]$ as the system size $M \propto N$ increases. In the case of sf/ps networks $\overline{\lambda^{1+\beta}}= \bar{\lambda}^{1+\beta}$ is finite and independent of $M$, leading to
\begin{equation}
-{\rm ln} [1- r_{u}(i,j)] \sim M^{-\beta}.
\label{eq:scaling_ps}
\end{equation}
In the case of sf/sf networks, $\overline{\lambda^{1+n\beta}}$ is given by Eq.~(\ref{eq:app:moment1}) and results in
\begin{equation}
-{\rm ln} [1- r_{u}(i,j)] \sim M^{-\frac{(\gamma-2)(1+\beta)}{\gamma-1}}.
\label{eq:scaling_sf}
\end{equation}
As can be seen from Eqs.~(\ref{eq:scaling_ps}) and (\ref{eq:scaling_sf}), the connection probability $r_{u}(i,j)$ of top nodes in the one-mode projection decreases as $M$ increases, and scales as $M^{-\eta}$, where exponent $\eta = \beta$ in sf/ps networks and $\eta=\frac{(\gamma-2)(1+\beta)}{\gamma-1}$ in sf/sf networks. This result implies that one-mode projections of sufficiently large bipartite networks are sparse.
However, we note that $\eta \to 0$ in sf/sf networks as the power law exponent of the bottom domain $\gamma \to 2$. This result implies that one-mode projections of finite sf/sf networks characterized by $\gamma$ close to $2$ and small $\beta$ values may be very dense and overinflated with fully connected subgraphs. This effect can render geometry inference using the one-mode projections inaccurate.

\end{widetext}

\bibliographystyle{unsrt}

\end{document}